\documentclass[final,3p,12pt,authoryear,a4paper]{elsarticle}
\usepackage{float}
\usepackage{amsmath,amsfonts,amssymb,amsthm,pxfonts,dsfont}
\usepackage{setspace}
\usepackage{natbib}
\usepackage[bookmarks, colorlinks, pdftitle={Partial Identification of Heteroskedastic Structural Vector Autoregressions: Theory and Bayesian Inference},pdfauthor={Helmut Luetkepohl, Fei Shang, Luis Uzeda, Tomasz Wozniak}]{hyperref}
\usepackage{graphicx}
\usepackage{xcolor}
\usepackage{bm}
\usepackage{geometry}
\usepackage{multirow}
\usepackage{booktabs}
\usepackage{longtable,lscape,ltcaption}
\usepackage{rotating}
\usepackage{soul}
\usepackage[british]{babel}
\usepackage[normalem]{ulem}
\usepackage{ragged2e}
\usepackage{float}
\usepackage{amsthm} 
\usepackage{pdfpages}
\usepackage{placeins}
\numberwithin{equation}{section}

\newtheorem{thm}{Theorem}[section]
\newtheorem{lm}{Lemma}[section]
\newtheorem{cl}{Corollary}[section]

\usepackage{etoolbox}

\makeatletter
\g@addto@macro\appendix{%
}
\makeatother

\emergencystretch=\maxdimen
\DeclareMathOperator{\diag}{diag}

\begin{document}

\biboptions{longnamesfirst}
\newtheorem{prop}{Proposition}
\newdefinition{df}{Definition}
\newdefinition{ex}{Example}
\newdefinition{as}{Assumption}
\newdefinition{pr}{Property}
\newdefinition{rs}{Restriction}
\newdefinition{al}{Algorithm}

\journal{}

\begin{frontmatter}


\title{Partial Identification of Structural Vector Autoregressions with Non-Centred Stochastic Volatility}

\author[fub]{Helmut L\"utkepohl}
\author[us]{Fei Shang}
\author[boc]{Luis Uzeda\corref{cor2}}
\author[um]{Tomasz Wo\'zniak\corref{cor1}}

\cortext[cor2]{The views expressed in this paper are those of the authors and do not necessarily reflect the position of the Bank of Canada.}
\cortext[cor1]{Corresponding author. \emph{Email address:} \href{mailto:tomasz.wozniak@unimelb.edu.au}{tomasz.wozniak@unimelb.edu.au}.}

\address[fub]{Freie Universit\"at Berlin \& DIW Berlin, Freie Universität Berlin, Boltzmannstr. 20, 14195 Berlin, Germany}
\address[us]{Guangdong University of Foreign Studies, Xiaoguwei, Panyu District, Guangzhou 510006, China}
\address[boc]{Bank of Canada, 234 Wellington St. W, Ottawa, ON K1A 0G9}
\address[um]{University of Melbourne, 111 Barry St., 3053 Carlton, VIC, Australia}

\begin{abstract}
\noindent We consider structural vector autoregressions that are identified through stochastic 
volatility under Bayesian estimation. Three contributions emerge from our exercise. 
First, we show that a non-centred parameterization of stochastic volatility yields a 
marginal prior for the conditional variances of structural shocks that is centred on 
homoskedasticity, with strong shrinkage and heavy tails---unlike the common centred 
parameterization. This feature makes it well suited for assessing partial 
identification of any shock of interest. Second, Monte Carlo experiments on small and large 
systems indicate that the non-centred setup estimates structural parameters 
more precisely and normalizes conditional variances efficiently. Third, revisiting prominent fiscal structural vector autoregressions, we show how the non-centred approach identifies tax shocks that are consistent with estimates reported in the literature.
\end{abstract}


\end{frontmatter}

\newpage
\section{Introduction}\label{sec:intro}

\noindent This paper examines the partial identification of a structural shock in structural vector autoregression models with stochastic volatility (SVAR-SV). Following \cite{Rubio-Ramirez2010}, we define partial identification of a structural shock as the case in which the parameters of its associated equation within the system are globally identified, up to a sign normalization as in \cite{wz2003norm}. Partial identification is particularly important in empirical SVAR applications, where attention often centers on a subset of shocks rather than the full set, or in cases where identification of all shocks is not attainable \citep[e.g.][]{lutkepohl_testing_2016}. Common examples include efforts to identify monetary or fiscal policy shocks \citep[see, e.g.,][]{blanchard_empirical_2002,romerromer2004,mertens_reconciliation_2014,lewis_identifying_2021}. 

Stochastic volatility, in turn, has become a widely adopted feature in structural models. It is commonly used to improve forecast performance \citep[see, e.g.,][]{stock_watson_2007,Clark2011,Clark2015,uzeda_2022} and to study the time-varying nature of economic shocks \citep[see, e.g.,][]{Primiceri2005,CogleySargent2005,baumeister_peersman_2013}. Despite its widespread use, stochastic volatility has not been extensively explored as a source of identification via heteroskedasticity in the spirit of the seminal contribution by \cite{Rigobon:03} with a notable exception being the recent study by \cite{Bertsche/Braun:18}.\footnote{We expand on the distinctions between our approach and theirs later. Also, to be clear, by stochastic volatility we refer to modeling time-varying volatility as a latent process in a state-space framework, noting that identification through heteroskedasticity has also been studied under alternative volatility specifications.} 

This paper contributes to fill this gap by combining stochastic volatility modeling with the partial identification of specific structural shocks of interest. To this end, we adopt a \emph{non-centred} parameterization for stochastic volatility models \citep[see, e.g.,][]{kastner2014ancillarity, Chan2016}. This approach differs from the standard \emph{centred} parameterization, which is widely used in various types of applied work with SVAR-SVs. In our setup, the \emph{non-centred} parameterization rewrites the law of motion for the log-volatility of each structural shock in a way that decouples this state variable from the scale parameters (i.e., the standard deviation of the innovations driving the log-volatilities). Similar transformations have been shown to improve computational efficiency in various classes of state-space models \citep[see, e.g.,][]{fruhwirth-schnatter_wagner_2010}. However, to date, no study has examined the benefits---or drawbacks---of this parameterization for identification through heteroskedasticity in SVAR-SV models.

In this sense, the paper makes three main contributions. First, we present new theoretical results on the identification of structural shocks through heteroskedasticity in Bayesian-estimated SVAR-SVs. Specifically, we formally characterize how the marginal prior on the conditional variances of structural shocks behaves under the non-centred parameterization and contrast it with the centred case. The key finding is that, compared to the centred specification, the non-centred approach yields a marginal prior centred at homoskedasticity, with strong shrinkage toward it and heavy tails. This setup ensures normalization of conditional variances around value 1, rendering precise estimates of the structural parameters. As an ancillary result to our main theoretical contribution, we also note that, under the non-centred setup, the conditions for partial identification can be verified in practice. We further outline how the Savage--Dickey density ratio (SDDR) can be used to test whether a shock of interest can be identified through stochastic volatility, and describe the conditions under which this approach is valid.

Our second contribution is computational. We conduct a comprehensive Monte Carlo exercise to evaluate the performance of the proposed shrinkage prior in non-centred stochastic volatility models relative to their centred counterparts, across both small and large systems. The results show that parameters crucial for identifying structural shocks—namely, the conditional variances of shocks and the relevant rows of the impact matrix—are estimated more precisely under the non-centred parameterization. 

Because the above-discussed theoretical and computational considerations suggest potential implications for practical settings, our third contribution is empirical. Specifically, we revisit several well-known fiscal-policy SVARs from the literature \citep[see][]{blanchard_empirical_2002,mertens_reconciliation_2014,lewis_identifying_2021} to show that our framework can identify tax shocks through heteroskedasticity, yielding estimates that align closely with those reported in previous studies. Our empirical assessment relies on a Gibbs sampler using advanced techniques: the structural matrix sampler \citep{Waggoner2003}, row-by-row autoregressive slope sampling \citep{chan_large_2021}, the auxiliary mixture method \citep{Omori2007}, and ancillarity-sufficiency interweaving \citep{kastner2014ancillarity}, enabling efficient simulation smoothing under uncertain heteroskedasticity. All procedures are implemented in \textbf{C++} via the \textbf{R} package \textbf{bsvars} \citep{bsvars,Wozniak2024}, yielding substantial speedups.

A comment is in order regarding our first contribution. Given the substantial evidence of time-varying volatility in macroeconomic and financial variables, the fact that the marginal prior from the non-centred parameterization is centred at, and shrinks toward, a homoskedastic shock may seem counterintuitive. The key nuance is that a \emph{structural shock} can be homoskedastic even when every \emph{variable} in a SVAR is heteroskedastic \citep[see][]{lutkepohl_testing_2016}. This is because the conditional variance of the variables in a SVAR can be represented in terms of the conditional variance of the reduced-form VAR residuals, which are convolutions of the structural shocks. For example, if a given structural shock appears in the reduced-form residuals of all VAR equations, it can occur that only this single shock exhibits time-varying volatility, while inducing every variable in the system to be heteroskedastic. Thus, a prior centred at, and shrinking towards, homoskedasticity with heavy tails does not contradict pervasive heteroskedasticity in the data; rather, it reflects that a small number of shocks can drive most of the observed volatility across variables. This interpretation aligns with evidence of comovement in volatility in economic and financial time series documented in factor model-based studies \citep[e.g.,][]{engel_kozicki_1992,carriero_clark_marcellino_2016,castelnuovo_tuzcuoglu_uzeda_2025}.

Perhaps the paper closest to ours is \cite{Bertsche/Braun:18}. However, our approach departs from theirs in several important ways. First, unlike them, we adopt a Bayesian inferential framework, arguably the predominant approach for estimating SVAR-SVs in empirical work. Second, this inferential setup enables us to explore theoretical considerations regarding the marginal prior for the conditional variances that emerge under different stochastic-volatility parameterizations. Third, we address issues of scalability—particularly estimation accuracy—when implementing the proposed identification scheme in larger systems, an exercise that, to the best of our knowledge, is undertaken here for the first time. Our paper also contributes to the growing literature on identification through heteroskedasticity using alternative models and techniques: \cite{lutkepohl_testing_2016} test for a heteroskedastic rank based on the number of independent heteroskedastic processes in a GARCH framework; \cite{lanneGMM2021b} propose moment-based tests exploiting kurtosis properties of the structural shocks; \cite{lewis_identifying_2021} introduces a non-parametric approach and tests based on autocorrelations of squared residuals under the assumption of non-proportional changes in structural volatilities; \cite{lutkepohl2021testing} examine testing procedures for two-regime models when the timing of volatility changes is known; and \cite{LW2017} develop an SDDR-based identification test using regime-switching heteroskedasticity.

The remainder of the paper is organized as follows. Section~\ref{sec:SVAR-SV} discusses heteroskedastic SVARs and presents the two stochastic volatility parameterizations considered. Section~\ref{sec:SVprior} characterizes the marginal prior for the conditional variances implied by these parameterizations. Section~\ref{sec:SDDR} outlines the conditions for applying the SDDR-based identification test in our framework. Section~\ref{sec:mc} presents two Monte Carlo studies to assess the estimation accuracy of the non-centred approach. Section~\ref{sec:empirical} reports our empirical results. Section~\ref{sec:Conclusion} concludes. A complementary theorem and its proof is stated in the Appendix, whereas technical details and additional results are given in the Supplementary Materials.

\subsection*{Notation}

\noindent The following notation applies to the main text and technical appendix: $\mathbf{y}$ denotes
the available data, $\mathbf{I}_N$ is the identity matrix of order $N$, $\mathbf{0}_{N\times N}$ and $\boldsymbol\imath_N$
are a matrix of zeros and a vector of ones of the indicated dimensions, respectively, the operator $\diag(\cdot)$ puts the vector provided as its argument on the main diagonal of a diagonal matrix, the indicator function $\mathcal{I}(\cdot)$ takes the value of 1 if the condition provided as the argument holds and 0 otherwise, $\otimes$ denotes the Kronecker product of matrices.
$A\setminus B$ defines the set with all elements of the set $A$
that are not in the set $B$.
$\Gamma(\cdot)$ denotes the gamma function, and $K_n(\cdot)$ denotes the modified Bessel function of the second kind. The following notation is used for statistical distributions: $\mathcal{N}$ stands for a univariate normal and $\mathcal{N}_N$ stands for the $N$-variate normal distribution. $\mathcal{NP}$ stands for a univariate normal product while $\log\mathcal{NP}$ for the univariate log normal product distribution (to be defined
in Section \ref{sec:SVprior}). The gamma distribution is denoted by $\mathcal{G}$,
the inverted gamma 2 by $\mathcal{IG}2$, and the uniform distribution by $\mathcal{U}$. Unless specified otherwise, $n$ goes from 1 to $N$, $t$ goes from 1 to $T$, and  $s$ goes from 1 to $S$.


\section{Two approaches to parameterize stochastic volatility in SVARs}\label{sec:SVAR-SV}

\noindent We begin with the following reduced-form VAR model of order $p$:
\begin{equation}\label{eq:RFVAR}
\mathbf{y}_t = \mathbf{A}_1 \mathbf{y}_{t-1} + \dots + \mathbf{A}_p \mathbf{y}_{t-p} + \mathbf{A}_d \mathbf{d}_t + \mathbf{u}_t,
\end{equation}
where $\mathbf{y}_t$ is an $N$-dimensional vector of
observable time series variables, $\mathbf{A}_i$, $i=1,\dots,p$, are $N\times N$
autoregressive coefficient matrices, $\mathbf{d}_t$ is a $d\times 1$ vector containing deterministic terms such as the intercept, trend variables, or dummies, $\mathbf{A}_d$ is the corresponding $N\times d$ matrix of coefficients, and $\mathbf{u}_t = (u_{1.t},\dots,u_{N.t})'$ is
an $N$-dimensional, zero-mean, serially uncorrelated error term.

The structural form links the reduced-form innovations $\mathbf{u}_t$ to the structural shocks $\mathbf{w}_t$ via the $N\times N$ contemporaneous effects matrix $\mathbf{B}_0$:
\begin{equation}\label{eq:SVAR-B0}
\mathbf{B}_0 \mathbf{u}_t = \mathbf{w}_t,
\end{equation}
where the structural shocks are contemporaneously uncorrelated. Their (possibly time-varying) unconditional or conditional variances are collected in the diagonal matrix
\begin{equation}\label{eq:SVAR-LAMBDA_t}
\boldsymbol{\Lambda}_t = \diag\left(\sigma_{1.t}^2,\dots,\sigma_{N.t}^2\right),
\end{equation}
so that the unconditional or conditional covariance matrices of the reduced-form errors are
$\boldsymbol{\Sigma}_t = \mathbb{E}[ \mathbf{u}_t\mathbf{u}_t' ]$ or $\mathbb{E}[ \mathbf{u}_t\mathbf{u}_t' \mid \mathbf{u}_{t-1}, \mathbf{u}_{t-2},\dots ]$, respectively.

It is well-known that the structural matrix $\mathbf{B}_0$ is not identified without additional restrictions. Theorem~A.1 in the Appendix provides general conditions for partial identification of individual rows of $\mathbf{B}_0$ in settings where heteroskedasticity supplies the identifying variation. The Theorem states that a structural shock is identified if a sequence of its appropriately scaled (conditional) variances is different from the corresponding sequences of (conditional) variances of all other shocks. These conditions are closely related to those presented by \cite{LW2017} in the context of Markov-switching heteroskedasticity. In particular, our setup generalizes \cite{LW2017} by allowing volatility changes at every point in time, rather than restricting them to a finite number of regimes. For conciseness, Theorem~A.1 is reported in the Appendix, but we will refer to it throughout the paper as needed.
   
We now shift our focus to heteroskedastic SVARs, where shocks are modeled and identified through stochastic volatility. Specifically, we examine two parameterizations of this process: the centred and non-centred approaches.

 \subsubsection*{The centred parameterization}

\noindent To fix ideas, we first present the more conventional (centred) parameterization for modeling stochastic volatility. In this setup, each diagonal element of $\bm{\Lambda}_t$ in \eqref{eq:SVAR-LAMBDA_t} is parameterized as follows:
\begin{align}
\sigma_{n.t}^2 &= \exp\left(\tilde{h}_{n.t}\right),\label{eq:condvarSVcentred}\\
\tilde{h}_{n.t}
&= \rho_n \tilde{h}_{n.t-1} + \tilde\upsilon_{n.t}~~~\text{s.t.}~~~\rho_n\in(-1,1),
\label{eq:SV-ARcentred}\\
\tilde\upsilon_{n.t}&
\sim\mathcal{N}\left(0,\omega_{n}^2\right)\qquad\text{for}~ n=1,\cdots,N~\text{and}~t=1,\cdots,T.\label{eq:SVlastcentred}
\end{align}
This model is often complemented by an  inverse-gamma prior for $\omega_n^2$ \citep[see][and others]{Primiceri2005,CogleySargent2005,Clark2015,carriero_clark_marcellino_2016,chan_large_2021}. Theorem~A.1 implies that identification depends on the sequence of conditional variances, $\{\sigma_{n.t}^2\}_{t=1}^{T}$, associated with a shock. The shock is identified if its conditional variance sequence is not proportional to any sequence of volatilities from another shock in the system. These conditions are satisfied under \eqref{eq:condvarSVcentred}--\eqref{eq:SVlastcentred}, as this representation assumes that the log-volatility state variable $\tilde{h}_{n.t}$ (and thus $\sigma_{n.t}^2$) evolves stochastically as a stationary autoregressive process.

In this setup, assessing identification through stochastic volatility reduces to testing whether $\omega_n^2=0$. If that is the case, the corresponding shock is homoskedastic and is identified only if all other $\omega_k^2\neq0,~n\neq k$. Conversely, if $\omega_n^2\neq0$, then by construction $\{\sigma_{n.t}^2\}_{t=1}^{T}$ is unique, and the condition for identification in Theorem~A.1 is satisfied. 

Nevertheless, implementing statistical tests for $\omega_n^2=0$ is challenging because zero lies at the boundary of the parameter space for $\omega_n^2$. Moreover, Bayesian methods that estimate stochastic volatility models under the centred parameterization typically use an inverse-gamma prior for $\omega_n^2$, whose domain is undefined at zero. To address these issues, we adopt the non-centred parameterization for $\sigma^2_{n.t}$, discussed next.

\subsubsection*{The non-centred parameterization}

\noindent To obtain the non-centred parameterization of $\sigma_{n.t}^2$, akin to \cite{kastner2014ancillarity} and \cite{Chan2016}, we first define $\tilde{h}_{n.t}=\omega_n h_{n.t}$. This transforms \eqref{eq:condvarSVcentred}--\eqref{eq:SVlastcentred} into
\begin{align}
\sigma_{n.t}^2 &= \exp\left(\omega_{n}{h}_{n.t}\right),\label{eq:condvarSV_noncentred}\\
{h}_{n.t}
&= \rho_n {h}_{n.t-1} + \upsilon_{n.t}~~~\text{s.t.}~~~\rho_n\in(-1,1),
\label{eq:SV-AR_noncentred}\\
\upsilon_{n.t}&
\sim\mathcal{N}\left(0,1\right)\qquad\text{for}~ n=1,\cdots,N~\text{and}~t=1,\cdots,T.\label{eq:SVlast_noncentred}
\end{align}
We further assume that $h_{n.0}=0$, ensuring $\sigma_{n.0}^2=1$ to satisfy the normalization condition in Theorem~A.1.

Importantly, the conditions in Theorem~A.1 hold under more general specifications for the covariance structure of the innovations, that is, when the innovations driving $h_{n.t}$ and $\tilde{h}_{n.t}$ are correlated, given that correlation does not imply proportionality of $\{\sigma_{n.t}^2\}_{t=1}^{T}$ across equations. Even perfect correlation of volatility shocks $v_{n.t}$ and $\tilde{v}_{n.t}$ does not imply proportional changes of the conditional variances $\sigma_{n.t}^2$ due to the non-linear transformation in Equations \eqref{eq:condvarSV_noncentred} and \eqref{eq:condvarSVcentred}, respectively. This proportionality arises only when the shocks are perfectly correlated, parameters $\rho_n$ and $\omega_n^2$ are equal to their counterparts across equations, and $\omega_n^2 \neq 0$, an extreme scenario excluded in our framework.

While there is a one-to-one mapping between the representations in \eqref{eq:condvarSVcentred}--\eqref{eq:SVlastcentred} and \eqref{eq:condvarSV_noncentred}--\eqref{eq:SVlast_noncentred}, they have markedly different implications for (i) the marginal prior distribution of $\sigma_{n.t}^2$ and (ii) the feasibility of proposing a statistical method to assess shock identification. These implications are elaborated in Sections \ref{sec:SVprior} and \ref{sec:SDDR}.

Two more comments are in order. First, unlike the centred parameterization, the non-centred representation for $\sigma^2_{n.t}$ is cast in terms of the standard deviation parameter $\omega_{n}$ instead of $\omega_{n}^2$. Given that $\omega_{n}$ can take both positive and negative values on a real line, the non-centred approach allows us to elicit priors for which $\omega_{n}$ is defined at zero. As shown later in Section \ref{sec:SVprior_params}, we propose a conditionally normal prior for $\omega_n$ centred at zero, which implies a gamma prior for $\omega_n^2$. The gamma prior allocates more mass near $\omega^2_n=0$ compared to the inverse-gamma prior typically used in the centred approach \citep[see][]{Chan2016}. Consequently, the non-centred approach enables stronger shrinkage toward homoskedasticity also providing normalisation of the conditional variances around value~1.

Second, it is easy to see from \eqref{eq:condvarSV_noncentred} that the likelihood function is invariant to sign at the $(\omega_{n},~h_{n.t})$ ordinate. This follows from the fact that both $(\omega_{n},~h_{n.t})$ and $(-\omega_{n},~-h_{n.t})$ yield the same value for $\sigma_{n.t}^2$. Consequently, the posterior for $\omega_{n}$ may be bimodal or unimodal around zero. Bimodality will only occur if $\omega_{n}$ (and, consequently $\omega_{n}^2$) is far from zero. Therefore, bimodality of the posterior for $\omega_{n}$ provides evidence that $\sigma_{n.t}^2\neq 0$, supporting the identification of structural shocks through stochastic volatility. For the purpose of identification, a bimodal (as opposed to unimodal) posterior for $\omega$ is desirable. We return to this point in the context of our empirical application in Section \ref{sec:empirical}.

Having distinguished two approaches to model $\sigma_{n.t}^2$, for the remainder of this paper, we adopt the non-centred parameterization unless explicitly stated otherwise.


\section{The marginal prior for $\sigma^2_{n.t}$} \label{sec:SVprior}

\noindent In the context of Bayesian estimation,
$\sigma^2_{n.t}$ can be characterized through its marginal prior distribution. However, as shown in the previous section,
$\sigma^2_{n.t}$ is a non-linear function of $h_{n.t}$, $\omega_{n}$, and $\rho$. This non-linearity complicates the assessment of how the choice of priors for these variables affects the marginal prior for $\sigma^2_{n.t}$.

To shed light on this matter, this section provides a detailed examination of the marginal prior for $\sigma_{n.t}^2$. This is achieved in two steps. First, the priors for the parameters that underlie $\sigma^2_{n.t}$, namely $\omega_{n}$ and $\rho$, are specified. Second, the proposed marginal prior for $\sigma^2_{n.t}$ is characterized, illustrating how it ensures centring and shrinkage towards a homoskedastic SVAR. In what follows, we focus on the characterization of a univariate prior for $\sigma^2_{n.t}$ and provide more general results for a multivariate distribution of $\{\sigma_{n.t}^2\}_{t=1}^{T}$ in Supplementary Materials.

\subsection{Priors for the parameters underlying $\sigma^2_{n.t}$}\label{sec:SVprior_params}
\noindent Once again, the parameters associated with the non-centred representation of $\sigma^2_{n.t}$ are $\omega_n$, the essential parameter in our setup that determines whether $\sigma^2_{n.t}$ changes over time, and $\rho_n$, the autoregressive parameter of the latent process $h_{n.t}$. We assume the following hierarchical prior structure for these parameters:
\begin{align}
\omega_n\mid \sigma_{\omega_n}^2 &\sim\mathcal{N}\left(0,{} \sigma_{\omega_n}^2\right),\label{eq:prioromega}\\
\sigma_{\omega_n}^2\mid \rho_n &\sim\mathcal{G}\left(\underline{S}, \underline{A}\right)\mathcal{I}\left(0<\sigma_{\omega_n}^2<1-\rho_n^2\right),\label{eq:priorsigmao}\\
\rho_n\mid \sigma_{\omega_n}^2 &\sim\mathcal{U}\left( - \sqrt{1-\sigma_{\omega_n}^2},{} \sqrt{1-\sigma_{\omega_n}^2} \right),\label{eq:priorrho} 
\end{align}
where $\sigma_{\omega_n}^2$ denotes the prior variance of $\omega_n$.

The prior specification for $\omega_n$ in \eqref{eq:prioromega} extends the one proposed by \cite{Chan2016}. Specifically, instead of fixing the prior variance as in \cite{Chan2016}, we adopt a hierarchical prior in which $\sigma_{\omega_n}^2$ follows the gamma distribution stated in \eqref{eq:priorsigmao}. Consequently, our specification allows for the estimation of $\sigma_{\omega_n}^2$, making the prior for $\omega_n$ less dependent on arbitrary choices. Moreover, based on the results of \cite{bitto2019achieving} and \cite{cadonna2020triple}, marginalizing the prior for $\omega_n$ over $\sigma_{\omega_n}^2$ yields a prior that combines extreme shrinkage towards homoskedasticity with heavy tails. The latter accommodates heteroskedasticity when it arises from strong data signals.

We complement the three priors above with the following three restrictions:
\begin{align}
\frac{\sigma_{\omega_n}^2}{1-\rho_n^2} &\leq 1,\label{eq:restrictionrhosigma}\\
\underline{A}&> 0.5,\label{eq:restrictiona}\\
|\rho_n| &<1.\label{eq:restrictionrho}
\end{align}
Restriction \eqref{eq:restrictionrhosigma} ensures the desired level of centring and shrinkage in our proposed marginal prior for $\sigma^2_{n.t}$, which we show formally in Section~\ref{ssec:marginal}. It restricts the prior variances from Proposition~\ref{prop:prioruni} presented below in the limit  $\lim\limits_{t\to\infty} \sigma_{\omega_n}^2\frac{1-\rho_n^{2t}}{1-\rho_n^2} = \sigma_{\omega_n}^2/\left(1-\rho_n^2\right)$ to ensure that the condition holds for variances at all periods $t$. The restriction in \eqref{eq:restrictiona} determines the marginal prior for $\omega_n$, making it particularly suitable for our setup. 
Restriction \eqref{eq:restrictionrho} is standard and ensures that $h_{n.t}$ in \eqref{eq:SV-AR_noncentred} is stationary. 
Additionally, Restrictions \eqref{eq:restrictionrho} and \eqref{eq:restrictionrhosigma} determine the bounds for $\rho_n$ as expressed in the uniform prior for $\rho_n$ in \eqref{eq:priorrho}. Similarly, the truncation of the gamma prior for $\sigma_{\omega_n}^2$ stated in \eqref{eq:priorsigmao} arises from Restriction~\eqref{eq:restrictionrhosigma}. We provide further elaboration on these restrictions later in this section. 

\subsection{Characterizing the marginal prior for $\sigma_{n.t}^2$}\label{ssec:marginal}

\noindent This section provides a detailed description of the marginal prior for $\sigma_{n.t}^2$. As discussed in the Introduction, a key feature of our prior setup for $\sigma^2_{n.t}$
is to ensure that this prior is not only centred on the hypothesis of a homoskedastic SVAR but also provides shrinkage toward it. In this regard, Definitions \ref{df:NP} and \ref{df:logNP}, along with Proposition \ref{prop:prioruni}, presented below, will be instrumental in structuring the approach to achieving these objectives.
\begin{df}\label{df:NP}
{\textbf{Normal product distribution}}

\noindent Let $x$ and $y$ denote two independent zero-mean normally distributed random variables with variances $\sigma_x^2$ and $\sigma_y^2$, respectively. Then,  random variable $z=xy$ follows the normal product distribution with zero mean and variance $\sigma_z^2=\sigma_{x}^2\sigma_{y}^2$, denoted by
$z\sim\mathcal{NP}\left(\sigma_z^2\right)$,
and density function given by
$\frac{1}{\pi\sqrt{\sigma_z^2}}K_0\left(\frac{| z|}{\sqrt{\sigma_z^2}} \right)$. \hfill $\Box$
\end{df}
The normal product distribution is known in the statistical literature. We state it
here to clarify our notation. However, the following distribution is new and its density function is obtained by a change of variables.
\begin{df}\label{df:logNP}
{\textbf{Log-normal product distribution}}

\noindent Let random variable $z$ follow the normal product distribution with variance $\sigma_z^2$. Then, random variable $q=\exp(z)$ follows the log-normal product distribution, denoted 
$q\sim\log\mathcal{NP}\left(\sigma_z^2\right)$,
with density given by:
$\frac{1}{\pi\sqrt{\sigma_z^2}}\frac{1}{q}K_0\left(\frac{| \log q|}{\sqrt{\sigma_z^2}} \right)$. \hfill$\Box$
\end{df}
Based on the results from Definitions \ref{df:NP} and \ref{df:logNP}, we can state Proposition \ref{prop:prioruni}:
\linebreak
\begin{prop}{\textbf{Auxiliary results on conditional distributions for $h_{n.t}$ and $\sigma^2_{n.t}$}}\label{prop:prioruni}
\noindent Given the prior specification from Equations \eqref{eq:condvarSV_noncentred}--\eqref{eq:SVlastcentred} and \eqref{eq:prioromega}--\eqref{eq:restrictiona}, the marginal priors for the latent process $h_{n.t}$, log-conditional variances $\log\sigma_{n.t}^2=\omega_nh_{n.t}$, and conditional variances $\sigma_{n.t}^2=\exp(\omega_nh_{n.t})$ are given by the following normal, normal product, and log normal product distributions:
\begin{description}
\item[(a)] $h_{n.t}\mid\rho_n\sim\mathcal{N}\left(0,{} \frac{1-\rho_n^{2t}}{1-\rho_n^2}\right)$,
\item[(b)] $\log\sigma_{n.t}^2 \mid \rho_n, \sigma_{\omega_n}^2 \sim\mathcal{NP}\left(\sigma_{\omega_n}^2\frac{1-\rho_n^{2t}}{1-\rho_n^2}\right)$,
\item[(c)] $\sigma_{n.t}^2 \mid \rho_n, \sigma_{\omega_n}^2 \sim\log\mathcal{NP}\left(\sigma_{\omega_n}^2\frac{1-\rho_n^{2t}}{1-\rho_n^2}\right)$.
\end{description}
\end{prop}

\begin{proof}
\textbf{(a)} The result is based on the properties of a normal compound distribution that facilitates the integration of $\int p(h_{n.t}, h_{n.t-1}, \dots, h_{n.1})d(h_{n.t-1}, \dots, h_{n.1})$, where the joint distribution under the integral is constructed from the conditional distributions $h_{n.t}\mid h_{n.t-1}, \dots, h_{n.1} \sim\mathcal{N}(\rho_n h_{n.t-1},1)$ and using $h_{n.0}=0$. \textbf{(b)} The result is obtained directly by applying Definition \ref{df:NP}, the result \textbf{(a)} and the prior in Expression \eqref{eq:prioromega}. Point \textbf{(c)} is obtained as a straightforward consequence of the first two results and Definition~\ref{df:logNP}.
\end{proof}

The introduced results facilitate centring and shrinking our prior for $\sigma_{n.t}^2$ toward a homoskedastic SVAR, which is useful because it ensures that evidence supporting heteroskedasticity -- and, consequently, the identification of a shock -- must come from the data. Moreover, it provides an alternative strategy for normalizing SVARs that does not rely on common approaches, such as setting the diagonal elements of $\mathbf{B}_0$ to one or imposing that the expected value of $\sigma^2_{n.t}$ equals one. Both of these approaches complicate the derivation of an efficient Bayesian estimation algorithm. In what follows, we discuss how we achieve centring and shrinking of our prior for $\sigma_{n.t}^2$.

To centre the prior for $\sigma^2_{n.t}$ around the hypothesis of homoskedasticity, we must ensure that the log-normal product distribution characterizing $\sigma^2_{n.t}$ has a single pole at the value~1. This follows directly from two points: (i) the normalization condition in Theorem~A.1, which sets $\sigma^2_{n.0}=1$, and (ii) the fact that homoskedasticity in our setup corresponds to setting $\omega_n=0$, which implies $\sigma^2_{n.t}=\exp(\omega_n h_{n.t})=1$, as discussed in Section \ref{sec:SVAR-SV}. Property~\ref{pr:onemode}, presented below, establishes when the log-normal product for $\sigma^2_{n.t}$ is proper and has a single pole at 1.

\begin{pr}\label{pr:onemode}\textbf{Single pole of log-normal product distribution at point 1}

\noindent The log-normal product distribution from Definition~\ref{df:logNP} has a single pole at point 1 when its variance satisfies
$\sigma_z^2\leq 1$.
In this case, the value of the density function approaches infinity when its argument, $q$, goes to 1 and approaches 0 when $q$ goes to 0 from the right. If $\sigma_z^2 > 1$, this distribution has an additional pole at 0, hence approaching infinity as $q$ goes to either 0 or 1.
\end{pr}

\begin{figure}[t!]
\begin{center}
\caption{Densities of the log-normal product distribution for various values of the scale parameter.}

\smallskip
\includegraphics[scale=0.7]{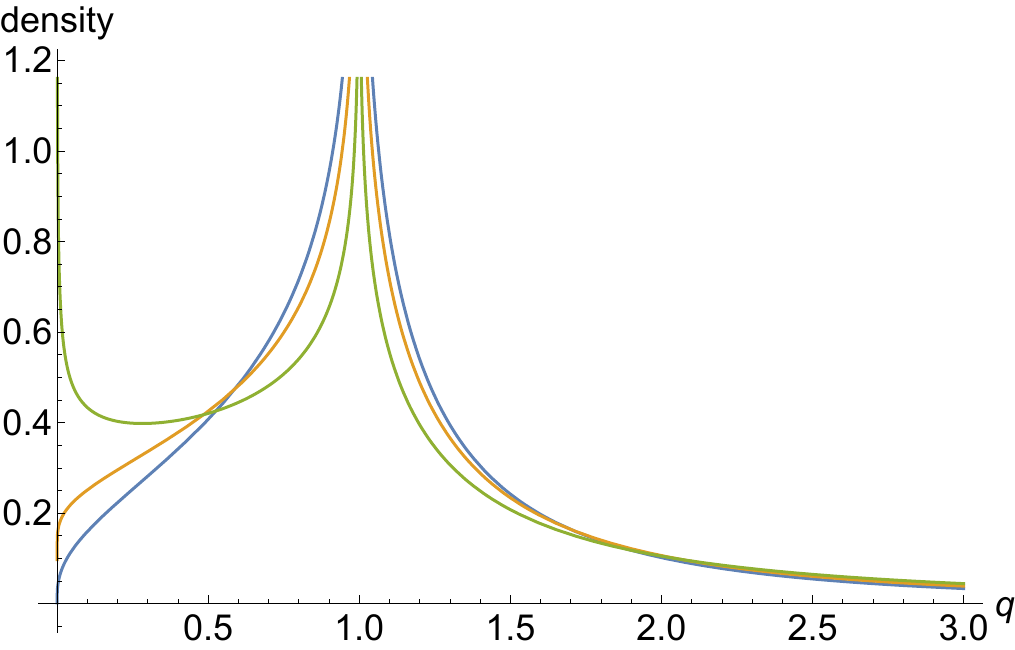}
\label{fig:logNP}
\end{center}\footnotesize

Note: The blue, orange, and green lines correspond to the densities for the values of the scale parameter $\sigma_z^2$ equal to 0.8, 1, and 1.5, respectively.
\end{figure}

Figure~\ref{fig:logNP} illustrates Property~\ref{pr:onemode}. A few points are worth highlighting. First, note that the condition for a single pole at one was stated in \eqref{eq:restrictionrhosigma}, which denotes a restriction on the variance of the prior limiting distribution for $\sigma^2_{n.t}$, as characterized in Proposition \ref{prop:prioruni}. This restriction ensures that the inequality in \eqref{eq:restrictionrhosigma} holds for all $t$. Second, the single-pole-at-one condition
implies a strong concentration of the prior probability mass for $\sigma^2_{n.t}$ at the value corresponding to the homoskedasticity of the structural shocks. This prior is equation invariant and, thus, it supports our claim that, at the prior mode, the SVAR model is not identified through heteroskedasticity.

The prior shrinkage for $\sigma^2_{n.t}$ is also achieved through Property~\ref{pr:onemode} via the inequality restriction in \eqref{eq:restrictionrhosigma}. Specifically, this restriction prevents the prior probability mass for $\sigma^2_{n.t}$ from being distributed more evenly over the interval from 0 to 1, as would occur in the presence of an additional pole at zero as for the density plotted in green in Figure~\ref{fig:logNP}.

\begin{figure}[t!]
  \centering
  \caption{The marginal priors for $\sigma^2_{n.t}$ and $\log(\sigma^2_{n.t})$ in their centred and non-centred parameterizations}
  
  \vspace{0.2cm}
  \begin{minipage}{0.48\textwidth}
    \centering
    $p\left(\sigma^2_{n.t}\right)$\\
    \includegraphics[width=\linewidth]{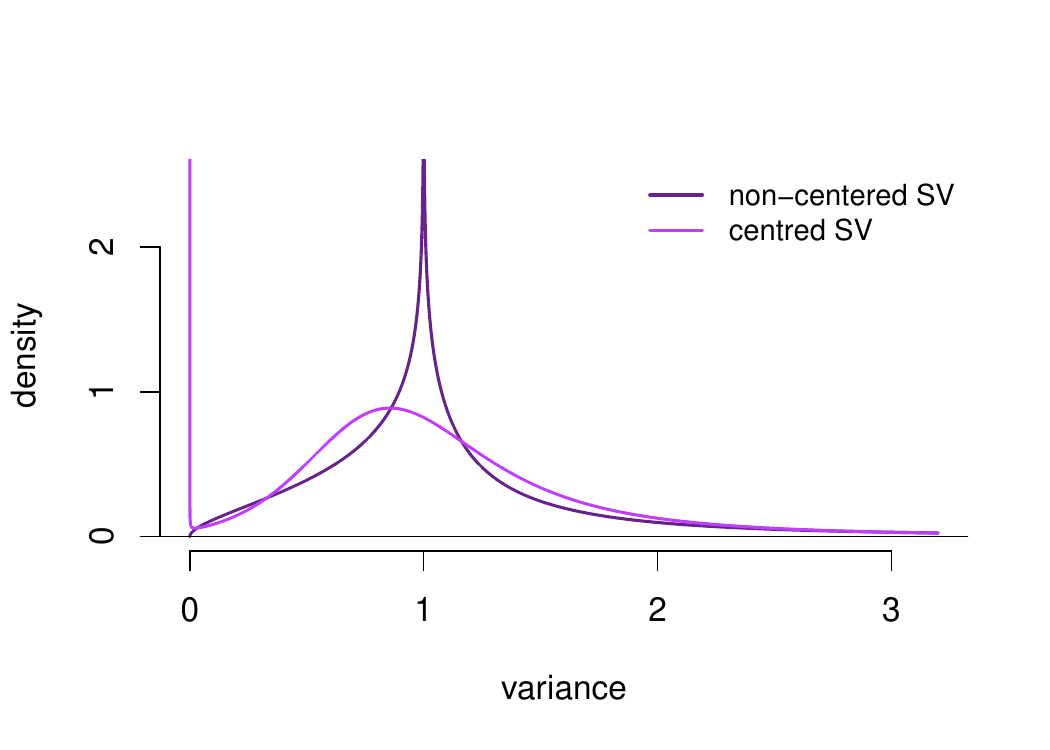}
  \end{minipage}
  \hfill
  \begin{minipage}{0.48\textwidth}
    \centering
    $p\left(\log(\sigma^2_{n.t})\right)$\\
    \includegraphics[width=\linewidth]{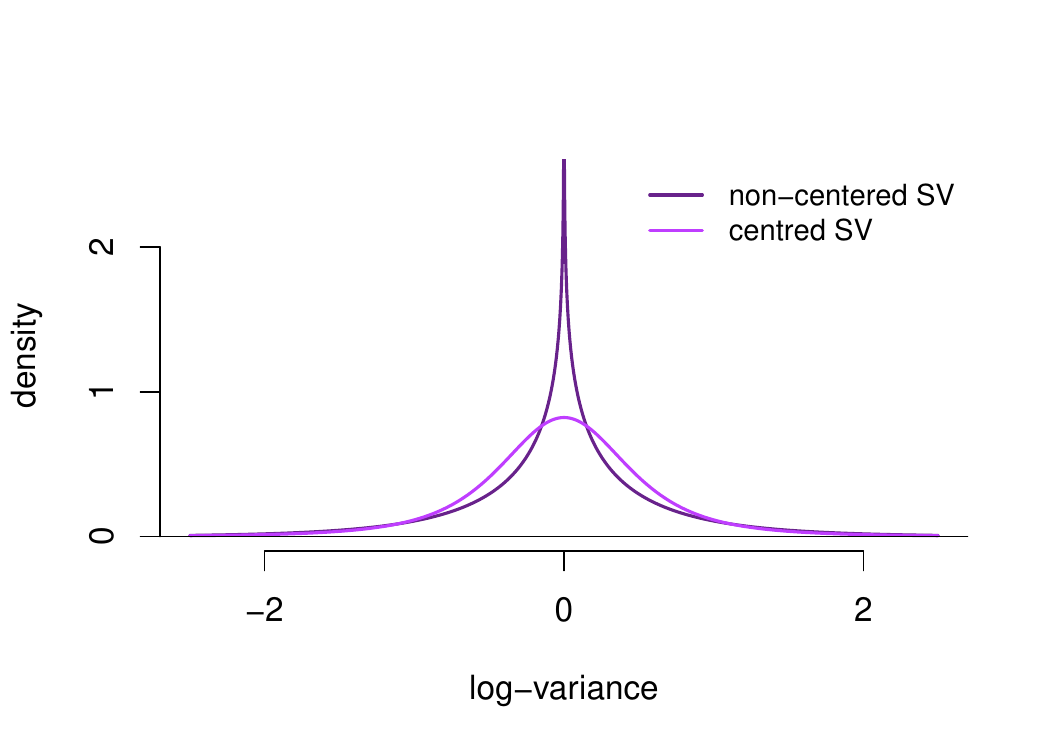}
  \end{minipage}
  
  \vspace{0.3cm} 
  \parbox{0.85\textwidth}{\footnotesize\justifying
    \noindent Note: The densities above correspond to the non-centred and centred approaches to model $\sigma^2_{n.t}$, as discussed in Section \ref{sec:SVAR-SV}.
  }

  \label{fig:distributions}
\end{figure}

The centering and shrinkage effects resulting from our prior setup are more evident in Figure~\ref{fig:distributions}, which compares the marginal prior distributions for $\sigma_{n.t}^2$ and $\log(\sigma_{n.t}^2)$ based on their centred and non-centred parameterizations.\footnote{The marginal priors in Figure \ref{fig:distributions} are computed using the numerical integration of \cite{Gelfand1990a} in two steps. In the first one for the non-centred parameterization, a sample of $S$ draws is obtained from the prior distributions, denoted by $\left\{\rho_n^{(s)},\sigma_{\omega_n}^{2(s)}\right\}_{s=1}^{S}$. In the second step, the marginal prior ordinates at pre-specified points, denoted by $\varsigma_g$ for $g=1,\dots,G$, are each computed by $\widehat{p}\left(\sigma_{n.t}^2=\varsigma_g\right) = S^{-1}\sum_{s=1}^{S}p\left(\sigma_{n.t}^2=\varsigma_g\mid\rho_n^{(s)},\sigma_{\omega_n}^{2(s)} \right)$. Appropriate modifications reflecting the prior assumptions are made for the centred parameterization computations.}

Notably, the marginal prior for $\sigma_{n.t}^2$ in the non-centred case inherits the properties from the log-normal product conditional prior for $\sigma_{n.t}^2$ given $\rho_n$ and $\sigma_{\omega_n}^{2}$, stated in Proposition \ref{prop:prioruni}\textbf{(c)} with a less-than-one restriction on its variance from Expression~\eqref{eq:restrictionrhosigma}. These properties are convergence to value zero when the conditional variance goes to zero from the right, a pole at 1, strong shrinkage toward the prior mode, and heavy tails. 

In contrast, in the centred case, the marginal prior for $\sigma^2_{n.t}$ has the properties of the log-Student-t distribution revisited in Supplementary Materials, that is, less concentration around the hypothesis of homoskedasticity, a mode at one, and a pole at zero. This log-Student-t distribution arises from the inverse gamma prior for $\omega^2_n$ featured by the centred parameterization and the exponential transformation of the log-volatility to conditional variance. In summary, the prior distribution for $\sigma_{n.t}^2$ in the centred parameterization favours heteroskedasticity, implies shock identification even in the absence of time-varying volatility, and does not support the normalization of the conditional variances at one.


\section{Bayesian Verification of Identification}\label{sec:SDDR}

\noindent Recall from Section \ref{sec:SVAR-SV} that in our non-centred setup, identifying a given shock $n$ thorough heteroskedasticity involves assessing the restriction $\omega_n = 0$. If this restriction holds, then $h_{n.t}=0$ and $\sigma_{n.t}^2=1$ for all $t$, which corresponds to a homoskedastic shock. Conversely, if $\omega_n\neq0$, then the conditions in Theorem~A.1 ensure shock identification through heteroskedasticity.

To assess $\omega_n = 0$, we adopt a Bayesian approach by comparing the fit of homoskedastic and a partially heteroskedastic SVAR using the Bayes factor. To compute the Bayes factor, we use the SDDR approach, which defines the Bayes factor as
\begin{equation}\label{eq:sddr}
BF_{homosk} = \frac{p(\omega_n=0|\mathbf{y})}{p(\omega_n=0)}.
\end{equation}
The numerator in \eqref{eq:sddr} is computed using numerical integration methods based on the estimator proposed by \cite{Gelfand1990a}. This approach only requires the full conditional posterior distribution of $\omega_n$ to be known up to its probability density function, which is normal, and the posterior draws from the unrestricted model ($\omega_n\neq0$). Consequently, computation of \eqref{eq:sddr} requires estimation only under the unrestricted, that is, heteroskedastic SVAR. Supplementary Materials provide a detailed description of the evaluation of the marginal posterior density $p(\omega_n=0|\mathbf{y})$.

The denominator $p(\omega_n=0)$ involves the marginal prior, which is obtained by integrating out $\sigma_{\omega_n}^2$ from the hierarchical-prior structure of $\omega_n$, discussed in Section \ref{sec:SVprior_params}, where $\omega_n|\sigma_{\omega_n}^2\sim\mathcal{N}(0,~\sigma_{\omega_n}^2)$ and $\sigma_{\omega_n}^2 \sim\mathcal{G}\left(\underline{S},{}\underline{A}\right)$. Proposition \ref{pr:marginalomega} formalizes this marginal prior as follows:

\begin{prop}\label{pr:marginalomega}\textbf{Density of the marginal prior $p(\omega_n)$}
\noindent The marginal prior density function for parameter $\omega_n$ obtained by marginalizing the joint prior distribution $p(\omega_n,\sigma_{\omega_n}^2)$ over $\sigma_{\omega_n}^2$,
$p\left(\omega_n\right) = \int_0^\infty p\left(\omega_n\mid\sigma_{\omega_n}^2\right)p\left(\sigma_{\omega_n}^2\right) d\sigma_{\omega_n}^2$,
where the priors $p\left(\omega_n\mid\sigma_{\omega_n}^2\right)$ and $p(\sigma_{\omega_n}^2)$ are given by
Expressions~\eqref{eq:prioromega} and \eqref{eq:priorsigmao}, respectively. This yields
\begin{align}\label{eq:marginalomega}
p\left(\omega_n\right) =
\frac{\left|\omega_n\right|^{\underline{A}-\frac{1}{2}} K_{\underline{A}-\frac{1}{2}}\left(\sqrt{\frac{2}{\underline{S}}}|\omega_n|\right)}
{\sqrt{\pi}\left(\sqrt{2}\right)^{\underline{A}-\frac{3}{2}} \Gamma\left(\underline{A}\right)\left(\sqrt{\underline{S}}\right)^{\underline{A}+\frac{1}{2}}  }.
\end{align}
\hfill$\Box$
\end{prop}
\begin{proof}
The integration proceeds by recognizing the constant and kernel and applies to the latter, which is facilitated using the normalizing constant of the generalized inverse Gaussian distribution provided by \cite{barndorff1997normal}.
\end{proof}

To compute the Bayes factor using the SDDR approach, it is crucial that the marginal prior $p(\omega_n)$ is bounded at $\omega_n=0$. Property \eqref{pr:bounded} establishes that the existence of this bound depends on the hyperparameter $\underline{A}$:

\begin{pr}\label{pr:bounded}\textbf{Upper bound of the marginal prior density for $\omega_n$} \citep[see][Theorem 2]{cadonna2020triple}.
\begin{align}
\lim\limits_{\omega_n\to 0} p\left(\omega_n\right) =
\left\{\begin{array}{cl}
\infty &\text{ for } 0<\underline{A}\leq 0.5,\\
\displaystyle
\frac{1}{\sqrt{2\pi \underline{S}}\left(\underline{A}^2-\frac{1}{4}\right)} \frac{\Gamma\left(\underline{A}+\frac{3}{2}\right)}{\Gamma\left(\underline{A}\right)} &\text{ for } \underline{A}>0.5. \qquad
\end{array}\right.
\end{align}
\end{pr}

\noindent Thus, the marginal prior density $p(\omega_n)$ is bounded from above if $\underline{A}> 0.5$, as required by Restriction~\eqref{eq:restrictiona}. Accordingly, we set $\underline{A}=1$, reducing the gamma prior to an exponential distribution, consistent with the Bayesian Lasso prior considered by \cite{belmonte_hierarchical_2014}. Other choices are possible and are reviewed by \cite{cadonna2020triple}. Additionally, we set the hyper-parameter $\underline{S}=0.05$, ensuring that nearly all prior probability mass for $\sigma_{\omega_n}^2$ lies within the interval $(0,{}1)$.


\section{Two Monte Carlo studies}\label{sec:mc}

\noindent In light of our new methods, practitioners may wonder about our model's ability to normalize the system, effectively estimate the structural parameters, and verify structural shocks in finite samples under a misspecified variance process. To address these questions, we conduct two comparative Monte Carlo (MC) studies. The first investigates the estimation efficiency of key parameters for partial identification under heteroskedasticity, namely, selected rows of the structural matrix ($\mathbf{B}_0$) and conditional variances of the structural shocks ($\sigma_{n.t}^2$). The second assesses the performance of our verification procedure for shock identification in Section \ref{sec:SDDR} under different volatility processes. To implement these exercises, we rely on the posterior simulation algorithm, with details provided in Supplementary Materials.

\subsection{Estimation efficiency under centred and non-centred stochastic volatility}

\noindent We highlight that the centred parameterization may fail to normalize SVARs—that is, to distinguish between $\mathbf{B}_0$ and $\sigma^2_{n.t}$—due to the hump-shaped, as opposed to shrinking, prior for $\sigma^2_{n,t}$ centred around one, as illustrated in Figure~\ref{fig:distributions}. This could cause distortions in the parameter estimates of both $\mathbf{B}_0$ and $\sigma^2_{n.t}$. Therefore, using simulated data, we investigate whether the non-centred approach improves the estimation accuracy of $\mathbf{B}_0$ and $\sigma^2_{n.t}$, and consequently enhances system normalization.

Specifically, our first MC exercise simulates 100 artificial datasets from six different data-generating processes (DGPs), which vary by system dimension $N$ and sample length $T$. For each dataset, we estimate SVAR-SVs under both centred and non-centred parameterizations of $\sigma_{n,t}^2$ to assess estimation precision. We measure accuracy using root-mean-squared errors (RMSEs). The DGPs include systems with $N \in \{3, 10, 20\}$ variables, matching the dimensions considered in our empirical application in Section~\ref{sec:empirical} and related VAR studies. We set the sample sizes to $T \in \{260, 780\}$, corresponding to 65 years of quarterly and monthly data, respectively. To focus on identification, we use a simplified DGP that ignores the SVAR’s autoregressive coefficients and concentrates on the structural matrix $\mathbf{B}_0$ and the conditional variances $\sigma^2_{n.t}$. The DGP for this exercise is thus given by:
\begin{align}
\mathbf{B}_0\mathbf{y}_t &= \mathbf{w}_t,\quad \mathbf{w}_t\sim\mathcal{N}_N\left(\mathbf{0}_N,\diag\left(\boldsymbol\sigma_t^2\right)\right),\label{eq:dgpsstruc}\\
\sigma_{n.t}^2 &= \exp\left(0.5{h}_{n.t}\right), \quad h_{n.t} = 0.92h_{n.t-1} + v_{n.t}, \quad v_{n.t}\sim\mathcal{N}(0,0.25).\label{eq:dgpssv}
\end{align}
For a three-variable system, we set $\mathbf{B}_0$ to \citeauthor{blanchard_empirical_2002}'s estimates. We extend it to ten- and twenty-variable systems by filling the remaining diagonal elements with random draws from a gamma distribution with mean 100 and variance 1000, and by filling the lower-diagonal elements with random draws from a zero-mean normal distribution with variance 4. The stochastic volatility process in~\eqref{eq:dgpssv} reproduces the relatively high volatility persistence observed in macroeconomic aggregates and, with fixed parameters, represents both the centred and non-centred volatility specifications. We then estimate the structural models in~\eqref{eq:dgpsstruc} using both the centred and non-centred volatility formulations, maintaining similar priors across the two specifications.

\begin{table}[t!]
\caption{Relative RMSEs of non-centred vs. centred stochastic volatility specifications}
\begin{center}
\begin{tabular}{c|cc|cc}
\toprule
&\multicolumn{2}{c}{$\mathbf{B}_0$}&\multicolumn{2}{c}{$\sigma_{n.t}$}\\[1ex]
\midrule
$N$&\multicolumn{2}{c}{$T$}&\multicolumn{2}{c}{$T$}\\[1ex]
&$260$& $780$& $260$& $780$\\
\midrule
$3$&  0.883& 0.743& 0.907& 0.846\\
$10$& 0.934& 0.723& 0.948& 0.853\\
$20$& 0.947& 0.778& 0.952& 0.896\\
\bottomrule
\end{tabular}
\end{center}

\smallskip{\small
Note: The table reports the ratios of the overall root-mean-squared error for structural matrix, $\mathbf{B}_0$, and conditional standard deviations, $\sigma_{n.t}$, for $t=1,\dots,T$. We report root-mean-squared errors for the parameters from the first three equations in all the systems with $N=3$, 10, and 20 variables. The values of ratios less than 1 indicate that the parameters in our SVAR model with non-centred SV are estimated more precisely than in the model with centred SV.}
\label{tab:eff_rmse}
\end{table}%

Table~\ref{tab:eff_rmse} reports the relative RMSEs for $\mathbf{B}_0$ and $\sigma_{n.t}$ for the first three equations in all systems. We calculate the errors as the difference between the posterior mean estimates and the corresponding values of the data-generating processes. The RMSEs include all elements from the first three rows of the structural matrix and the conditional standard deviations for the first three equations across all periods. We compute the relative  RMSEs with the non-centred specification in the numerator and the centred specification in the denominator. Hence, values below one indicate superior estimation accuracy for the non-centred version.

The results for $\mathbf{B}_0$ show that the non-centred specification consistently outperforms the centred one in terms of RMSE. This difference matters because the structural parameters determine quantities such as impulse responses and forecast error variance decompositions. RMSE reductions for the non-centred specification range from about 5\% for the shorter sample ($T=260$) to around 22\% for the longer sample ($T=720$), and tend to decrease as the system dimension $N$ increases for a given sample size. Supplementary Materials report RMSE results for individual parameters in the top-left $3\times 3$ block of $\mathbf{B}_0$, showing that most parameters exhibit similar improvements across the different DGPs.

The results for $\sigma_{n.t}$ show similar patterns. Table~\ref{tab:eff_rmse} reports RMSE reductions ranging from about 5\% in the short sample ($T=260$) to roughly 10\% in the longer sample ($T=780$). These gains tend to increase slightly as the system dimension $N$ decreases for a given sample size. Supplementary Materials present additional results indicating that the improvements hold across all periods in the simulated data. Overall, these findings highlight the ability of the non-centred specification to more accurately normalize SVAR-SVs and support its empirical suitability in both smaller and larger models.

\subsection{Assessing the verification procedure for partial identification}

\noindent As discussed in Section~\ref{sec:SDDR}, verifying whether a shock is partially identified boils down to testing for homoskedasticity. To explore how well the SDDR procedure detects homoskedastic shocks, our second MC exercise evaluates its performance under two alternative prior specifications for $\omega_n$: (i) our hierarchical prior structure in \eqref{eq:prioromega}–\eqref{eq:priorrho}, and (ii) the zero-mean normal prior with fixed variance $\sigma_{\omega_n}^2 = 10$ proposed by \cite{Chan2016}. The latter violates the scaling restriction in \eqref{eq:restrictionrhosigma}, potentially affecting normalization.\footnote{We note, however, that the prior for $\omega_n$ in \cite{Chan2016} was implemented within a reduced-form setup, where normalization is less of a concern.} For this analysis, we implement the SDDR procedure using the non-centred parameterization of a system with stochastic volatility under both prior specifications.

We assess these priors by estimating the models on multiple artificially generated datasets. All DGPs in this exercise are bivariate and follow the structural equation~\eqref{eq:dgpsstruc}, with the structural matrix $\mathbf{B}_0$ chosen to reflect parameter estimates from our empirical application in Section~\ref{sec:empirical}. The only variation across DGPs is in the volatility process, for which we consider three alternative specifications:
\begin{description}
\item[SV:] following the specification in \eqref{eq:dgpssv}.
\item[GARCH:] where $\sigma_{n.t}^2 = 0.02 + 0.28u_{n.t-1}^2 + 0.7\sigma_{n.t-1}^2$ and $\sigma_{n.0}^2 = 1$.
\item[Markov switching heteroskedasticity (MSH):] where $\sigma_{n.t}^2 = \sigma_{n.s_t}^2$, $s_t$ is a two-state Markov process with transition probabilities $\mathbf{P}=\begin{bmatrix}0.98&0.02\\0.02&0.98\end{bmatrix}$, for $s_t=1$,
$\mathbf{w}_t\sim\mathcal{N}\left(\mathbf{0}_2,I_2\right)$ and
for $s_t=2$, $\mathbf{w}_t\sim\mathcal{N}\left(\mathbf{0}_2,\diag\left(20,10\right)\right)$.
\end{description}
Clearly, only the first volatility specification corresponds to our Bayesian verification of identification procedure in Section~\ref{sec:SDDR}, which implies that this procedure is violated when volatility changes follow either of the other two models. In other words, our second MC exercise provides a way to assess the robustness of our approach to misspecification.

\begin{table}[t!]
\begin{center}
\caption{Simulation results: rejection rates for homoskedasticity using our prior vs. \cite{Chan2016}}
\label{tab:rej}

\smallskip
\begin{tabular}{cp{3.65cm}|ccc|ccc}
\toprule
&&\multicolumn{3}{c|}{Our prior} & \multicolumn{3}{c}{\cite{Chan2016} prior}\\
\midrule
&&\multicolumn{3}{c|}{DGPs}&\multicolumn{3}{c}{DGPs}\\
\multirow{2}*{$T$}&homoskedastic shocks in each DGP &\multirow{2}*{SV}&\multirow{2}*{GARCH}&\multirow{2}*{MSH}&\multirow{2}*{SV}&\multirow{2}*{GARCH}&\multirow{2}*{MSH}\\
\midrule
\multicolumn{8}{c}{Panel A: $l$-value approach}\\[1ex]
\multirow{4}*{260}&shocks 1 \& 2 &0.00 & 0.00 & 0.00 & 0.00 & 0.00 & 0.00 \\
&shock 1 &0.01 & 0.01 & 0.00 & 0.02 & 0.00 & 0.01\\[1ex]
&shock 2 &0.57 & 0.31 & 0.19 & 0.55 & 0.27 & 0.30\\
&none &0.78 & 0.37 & 0.41 & 0.74 & 0.28 & 0.61\\[2ex]
\multirow{4}*{780}&shocks 1 \& 2 &0.00 & 0.00 & 0.00 & 0.00 & 0.00 & 0.00 \\
&shock 1 &0.00 & 0.00 & 0.02 & 0.01 & 0.00 & 0.01 \\[1ex]
&shock 2 &0.98 & 0.80 & 0.22 & 0.95 & 0.71 & 0.18 \\
&none &1.00 & 0.83 & 0.56 & 0.98 & 0.73 & 0.49 \\[1ex]
\midrule
\multicolumn{8}{c}{Panel B: $q$-value approach}\\[1ex]
\multirow{4}*{260}&shocks 1 \& 2 &0.05 & 0.05 & 0.05 & 0.05 & 0.05 & 0.05\\
&shock 1 &0.07 & 0.12 & 0.10 & 0.07 & 0.10 & 0.07\\[1ex]
&shock 2 &0.85 & 0.55 & 0.37 & 0.82 & 0.59 & 0.39\\
&none &0.95 & 0.68 & 0.78 & 0.96 & 0.69 & 0.78\\[2ex]
\multirow{4}*{780}&shocks 1 \& 2 &0.05 & 0.05 & 0.05 & 0.05 & 0.05 & 0.05 \\
&shock 1 &0.08 & 0.13 & 0.09 & 0.07 & 0.11 & 0.10\\[1ex]
&shock 2 &0.99 & 0.94 & 0.53 & 0.99 & 0.94 & 0.54\\
&none &1.00 & 0.97 & 0.91 & 1.00 & 0.98 & 0.90\\[1ex]
\bottomrule
\end{tabular}
\end{center}

\smallskip\footnotesize{Note: The table reports rejection rates for the hypothesis of homoskedasticity in the first shock, i.e., $\mathcal{H}_0: \omega_1=0$ using simulated data. The rates are calculated based on 100 realizations of DGPs each with the following characteristics: sample sizes: $T\in\{260,780\}$; volatility processes: SV, GARCH, MSH; homoskedastic shock arrangements: shocks 1 \& 2, shock 1, shock 2, none. For a homoskedastic shock, the variance is set to $\sigma_n^2 = 1$.}
\end{table}

For each volatility process, we generate data using four different 
scenarios: (1) both shocks are homoskedastic, (2) the first
shock is homoskedastic while the second shock is
heteroskedastic, (3) the first shock is heteroskedastic while 
the second shock is homoskedastic, (4) both shocks are
heteroskedastic. For homoskedastic shocks we set 
$\sigma_{n.t}^2=1$ for all $t$, 
while the heteroskedastic shocks are 
generated by the two different volatility models. 
Akin to the first MC exercise, we consider two sample sizes, $T \in \{260, 780\}$, and generate 100 simulated datasets for each of the four shock scenarios described above.

Table \ref{tab:rej} reports rejection rates for testing the homoskedasticity of the first shock. These rates are obtained using two strategies for constructing critical values, which we label $l$-value and $q$-value following \cite{BJ95} and \cite{Storey02}. In the $l$-value approach (Panel A), we apply a decision-theoretic rule and reject homoskedasticity if $BF_{homosk}<1$, where $BF_{homosk}$ is defined in Equation~\eqref{eq:sddr}; that is, rejection occurs when the posterior assigns more than 50\% probability to heteroskedasticity. In the $q$-value approach (Panel B), the critical value corresponds to the fifth percentile of the posterior odds ratio $BF_{homosk}$ computed under the null $\omega_1=0$. As a result, the rejection rate in the first row of Panel B is fixed at 0.05.

The rejection rates in Panel A of Table~\ref{tab:rej} indicate that our approach generally yields higher rejection power than the prior in \cite{Chan2016}, although the latter performs better for the Markov-switching DGP with the smaller sample size. Overall, both priors identify homoskedasticity of the first shock effectively and show strong performance in rejecting it when the first shock follows either SV or GARCH volatility. Rejection becomes more difficult when the volatility process is Markov-switching. Panel B confirms these results, with both priors delivering very similar rejection rates. Overall, our second MC exercise shows that the homoskedasticity verification method provides a reliable way to assess partial identification through heteroskedasticity. These results further support the use of the method in practice.\footnote{We acknowledge that several Bayesian and frequentist methods have been proposed for testing identification through heteroskedasticity in structural VARs, including \cite{LanneSaikkonen07}, \cite{lutkepohl_testing_2016}, \cite{lewis_identifying_2021}, and \cite{LW2017}. We ran simulations for all of these methods and report the results in Supplementary Materials. Since their null hypotheses differ from ours, they are not directly comparable. Moreover, some of these methods are tailored to specific volatility models and can perform well in such settings, but none consistently outperforms our approach across all scenarios. Each exhibits clear weaknesses in parts of our simulation design, so none emerges as a generally preferable alternative.}


\section{Empirical application: identifying tax shocks}\label{sec:empirical}

\noindent In this section, we illustrate the usefulness of the non-centred approach for identifying tax shocks. This empirical exercise is estimated using a posterior simulation algorithm, with details provided in Supplementary Materials.

When heteroskedasticity is used for identification in SVAR
analysis, the shocks are distinguished by their variances
or conditional variances. This approach provides
distinct shocks without economic labels and requires
some additional information to label the shocks.
Such information is sometimes available in the form
of specific shapes of the impulse responses associated with
a shock or a specific sign pattern of the impact effects of the
shocks.

To illustrate the methods
developed in the previous sections, we will consider a fiscal
SVAR model in which the unanticipated tax shock has been identified in different ways.
These alternative identification strategies include, for example,
\cite{blanchard_empirical_2002} (henceforth BP), who use
restrictions on the short-run effects of the shocks
and the instantaneous interactions of the variables to
identify their shocks, and by \cite{mountford2009} using sign
restrictions. Moreover,
\cite{mertens_reconciliation_2014} (henceforth MR), as revised by
\cite{ramey2016macroeconomic}, use an external
instrument, a narrative measure of the tax shock proposed by \cite{romer_macroeconomic_2010}.
Finally, \cite{lewis_identifying_2021} (henceforth LE)
uses heteroskedasticity and, hence, an approach in that
respect similar to ours. We use the MR model as our benchmark
to illustrate the use of our methodology for identifying
the tax shock through heteroskedasticity, and the narrative measure
by \cite{romer_macroeconomic_2010} to ensure a correct
labelling of the shocks.

\subsection{A simple fiscal SVAR}\label{ssec:sf}

\noindent MR specify a three-variable fiscal system including total tax revenue, denoted by $ttr$, government spendings, $gs$, and gross domestic product, $gdp$, and they express all the quarterly variables in real, log, per person terms. We will also
consider these three variables and investigate whether the
tax shock can be identified by our methodology.

\begin{figure}[t]
 \caption{Time series of three variables across three samples used for estimation}

\begin{center}
\includegraphics[width=\textwidth]{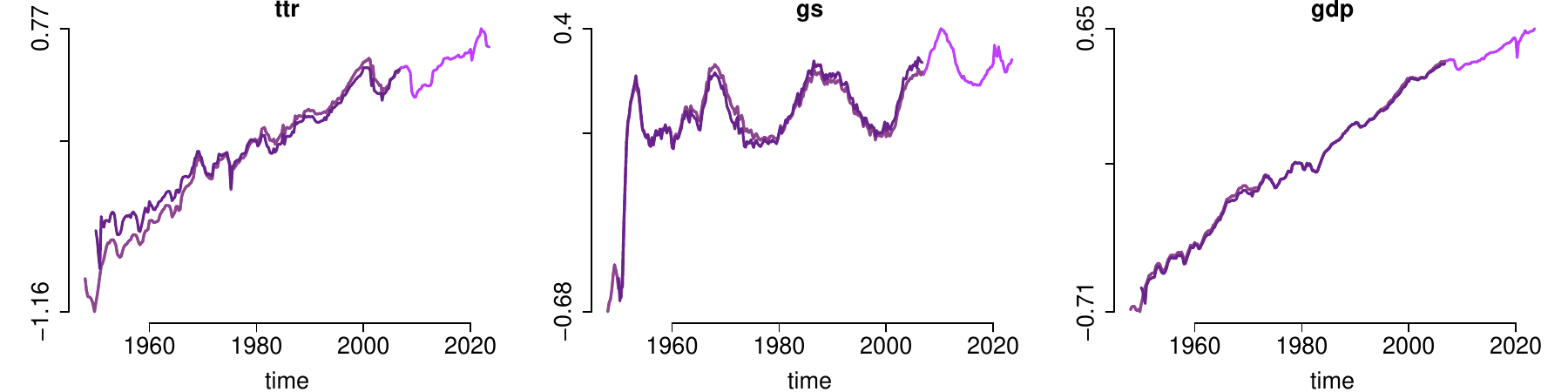}
\end{center}

\smallskip{\small
Note:  The figure plots three series for three samples:
the 2023-sample plotted in light pink includes observations from  1948Q1 to 2023Q3 ($T=303$), the 2006-sample plotted in darker pink is as the 2023-sample but finishes in  2006Q4 ($T=236$), the
MR-sample, plotted in purple, spans the period from 1950Q1 to
2006Q4 ($T=228$). The plotted series are standardized by subtracting from each series its first observation in~1980.}
\label{fig:DATA}
\end{figure}

In order to investigate identification through
heteroskedasticity in this fiscal system,
we use three alternative samples of different lengths and partly different values even for overlapping periods. They
are plotted in Figure~\ref{fig:DATA}, where it can be seen that
the series are different but similar in overlapping periods.
The shortest sample, hereafter MR-sample, uses the data from
MR and LE that is downloaded directly from Karel Mertens' website.\footnote{\label{fn:data}The spreadsheet is available at \href{https://karelmertenscom.files.wordpress.com/2017/09/jme2014_data.xls}{https://karelmertenscom.files.wordpress.com/2017/09/jme2014\_data.xls}}
Following the data construction described by MR, total tax revenue,
government spending, and gross domestic product, as well
as the GDP deflator are taken from NIPA Tables numbers 3.2,
3.9.5, 1.1.5, and 1.1.9, respectively, provided by the \cite{t32,t395,t115,t119}, and the population variable is
provided by \cite{francis2009measures}. This data spans
the period 1950Q1 to 2006Q4.

We extend the sample to the latest available observations in
2023Q3 with modifications in the population variable that is replaced by one matching \citeauthor{francis2009measures}'s (\textcolor{blue}{2009}) definition and provided by the \cite{pop}. Based on these variables we form two samples, both of which contain longer time series
than MR and
start in 1948Q1. One of these samples, hereafter the 2023-sample, ends in 2023Q3, and the other one,
hereafter the 2006-sample, ends in 2006Q4. Following MR,
we use a VAR(4) model with a~constant term,
a linear and a quadratic trend, and a dummy for 1975Q2
as deterministic terms.

\subsection{Verifying partial identification through heteroskedasticity}\label{ssec:ivh}

\noindent We base our structural analysis on model
(\ref{eq:SVAR-B0}). Hence, we have to sample from the
posterior of the structural $\mathbf{B}_0$ matrix,
which is not identified without further restrictions
if the shocks are homoskedastic. Even if the shocks are
identified, the
row ordering and row signs may change in different drawings
from the posterior if one does not take special precautions to prevent
that from happening. We, therefore, follow LE and reorder the rows and adjust
their signs such that each draw has the minimum distance to
the benchmark $\mathbf{B}_0$ matrix as in BP. More details on this procedure are provided in Supplementary Materials.\footnote{We also experiment with alternative benchmark estimates for $\mathbf{B}_0$. The results are virtually unchanged and are reported in Supplementary Materials.}
Hence, the shocks can be labeled along the lines of BP as an
unanticipated tax shock ($w_t^{ttr}$), a government
spending shock ($w_t^{gs}$), and an additional
shock ($w_t^{gdp}$) capturing unexpected changes in $gdp_t$
not caused by tax or spending shocks. We will label our
shocks accordingly
although it is, of course, not clear from the outset
if the shocks can be identified through heteroskedasticity
with our methodology. If they can, they may still differ
from those in BP and MR, in which case our labels may not be
meaningful. We will return to this issue later.

\begin{table}[t]
    \caption{Verification of partial identification of structural shocks through heteroskedasticity across samples}

    \begin{center}
    \begin{tabular}{ccccccc}
    \toprule
    Structural shocks &\multicolumn{2}{c}{2023-sample} &
    \multicolumn{2}{c}{2006-sample} &
    \multicolumn{2}{c}{MR-sample}\\
\midrule
$w_{t}^{ttr}$ & \textbf{-21.38} & [4.69] & -1.51 & [0.18] & 0.32 & [0.05] \\
  $w_{t}^{gs}$ & \textbf{-4.62} & [0.79] & -1.32 & [0.15] & 0.23 & [0.05] \\
  $w_{t}^{gdp}$ & \textbf{-63.39} & [6.43] & 0.50 & [0.03] & 0.39 & [0.03] \\
\bottomrule
    \end{tabular}
    \end{center}

\smallskip{\small
Note:  The table reports the log of the Bayes factors estimated via the log of SDDRs from Equation~\eqref{eq:sddr} together with numerical standard errors (NSEs) provided in brackets. Negative values provide evidence against homoskedasticity. Bold font numbers represent cases in which the evidence for heteroskedasticity is positive (values greater than 3 in absolute terms) or strong (greater than 20) on the scale of \cite{kassBayesFactors1995}. The NSEs are computed based on 30 subsamples of the original MCMC draws.
}
\label{tab:SDDR}
\end{table}

We next assess whether the labeled shocks are identified through heteroskedasticity.
Our main tool for that purpose is the SDDR
from Equation~\eqref{eq:sddr}. The SDDR values computed
for each of the three shocks individually using our three
data samples are reported in Table~\ref{tab:SDDR}.
For the 2023-sample, the evidence for heteroskedasticity
of all three structural shocks is strong according to the scale proposed by \cite{kassBayesFactors1995}. The values of the log Bayes factors shown in Table~\ref{tab:SDDR} indicate that the posterior mass in favour of heteroskedasticity exceeds 99\% for all the shocks. This result provides strong evidence for the identification of all three shocks through heteroskedasticity in the 2023-sample and is robust to many variations in the model prior specification. These variations include perturbations of the hyper-parameters that need to be fixed in our setup. We checked the conclusions for three values of each scale and shape of the prior distribution for $\omega_n$, as well as for three alternative setups for the hyper-parameters for each of the matrices $\mathbf{A}_i$ and $\mathbf{B}_0$. Each of these alternative setups included cases of stronger and weaker shrinkage than in our benchmark prior specification.

The evidence for the structural shocks to be identified through heteroskedasticity is much weaker in the 2006-sample.
Moreover, the log Bayes factors estimated by the log-SDDRs for the MR-sample are positive, implying that the posterior mass for homoskedasticity is greater than that for heteroskedasticity. The log-SDDRs are negative for the last two shocks in the 2006-sample, which includes eight more observations than the MR-sample from the volatile late 1940s. More specifically, in the 2006-sample,
the posterior probability of the heteroskedastic shock $w_{t}^{ttr}$ is 82\%. Obviously, in this case, the evidence for identification through heteroskedasticity of the first shock is limited
and it is even more limited for the other shocks.
These findings are also robust to the perturbations in the values of the prior hyper-parameters.

\begin{figure}[t]
 \caption{Marginal prior (solid line) and posterior (histograms)
densities of $\omega_{ttr}$, $\omega_{gs}$, and $\omega_{gdp}$ across samples}
\begin{center}
\begin{tabular}{p{2cm}p{5cm}p{5cm}p{5cm}}
&$\omega_{ttr}$ &$\omega_{gs}$ &$\omega_{gdp}$
\end{tabular}
\includegraphics[trim={0 0 0 1cm},clip,width=\textwidth]{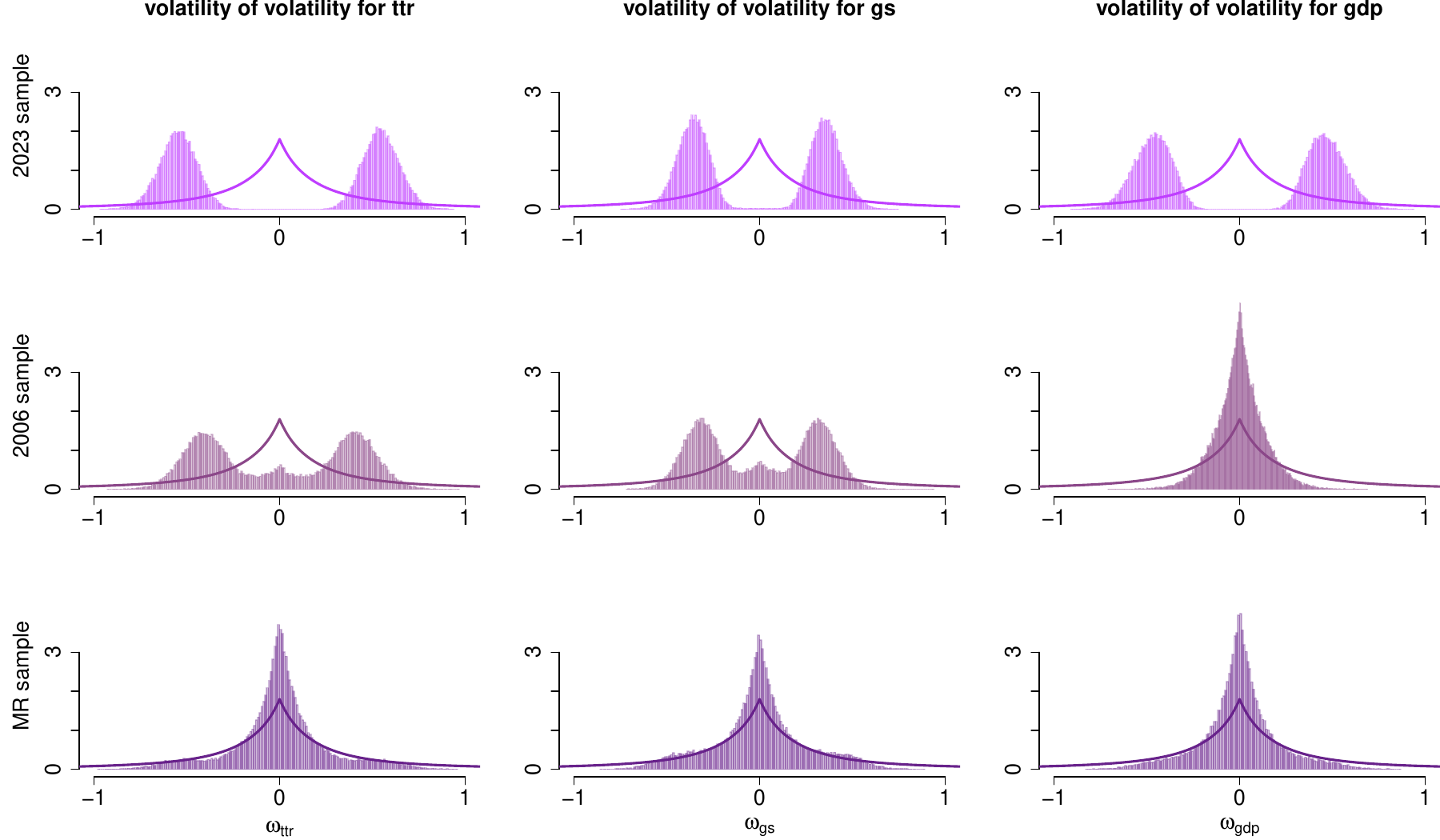}
\end{center}

\smallskip{\small
Note: The marginal prior density is estimated by numerical integration as in \cite{Gelfand1990a} using a grid of points from -1.1 to 1.1. They are the same for all samples and shocks. The marginal posterior densities are approximated using histograms. The ratio of these densities at point zero approximates the SDDR in Equation~\eqref{eq:sddr}. Posterior mass less concentrated than the prior mass about zero provides evidence against homoskedasticity.}
\label{fig:OMEGA}
\end{figure}

In Figure~\ref{fig:OMEGA}, we further illustrate how the SDDRs work by plotting the marginal prior versus the marginal posterior densities of $\omega_n$ associated with our three samples.
Based on the information from these plots, the SDDRs from Equation~\eqref{eq:sddr} can be approximated by the ratio of the marginal posterior ordinate at zero to that of the marginal prior density. The figures for the 2023-sample exhibit posterior mass concentrated away from the origin and the bi-modality discussed in Section~\ref{sec:SVAR-SV}, providing evidence against homoskedasticity. Instead,
the posterior mass for the 2006- and
MR-samples is concentrated about the hypothesis of homoskedasticity, often more than the prior, thus favouring
homoskedasticity.

Finally, we analyze the sequences of conditional variances
of the structural shocks that are required to be clearly distinct
for partial identification of the shocks to hold according to Theorem~A.1. We plot their posterior means together with 90\% highest posterior density (HPD) intervals in Figure~\ref{fig:HV}.\footnote{See Supplementary Materials for the centred parameterization version of these time-varying conditional variances.}
The conditional variances are visibly time varying for the 2023-sample. The conditional variances of the first shock are significantly different from 1 in six
periods in that sample, including the mid-70s and mid-80s, individual quarters in 2001, 2002, and 2003, and the first quarter of 2009. The variances of the second shock are
different from 1 in the first quarter of 1951 only, while
those of the third shock have HPD intervals not including 1 in 1950 and quarters 2 and 3 of 2020. The distinctive occurrence times of high volatility periods for the three shocks provide strong evidence for them to be different in these sequences, further supporting the identification through heteroskedasticity in this sample. In particular, this evidence supports our claim that the first shock is identified as its conditional variances evolve non-proportionally to those of other shocks.

\begin{figure}[t]
 \caption{Conditional variance of structural shocks across samples}

\begin{center}
\begin{tabular}{p{5cm}p{5cm}p{5.1cm}}
conditional variance of $w_t^{ttr}$ &conditional variance of $w_t^{gs}$ &conditional variance of $w_t^{gdp}$\\
\end{tabular}
\includegraphics[trim={0 0 0 1cm},clip,width=\textwidth]{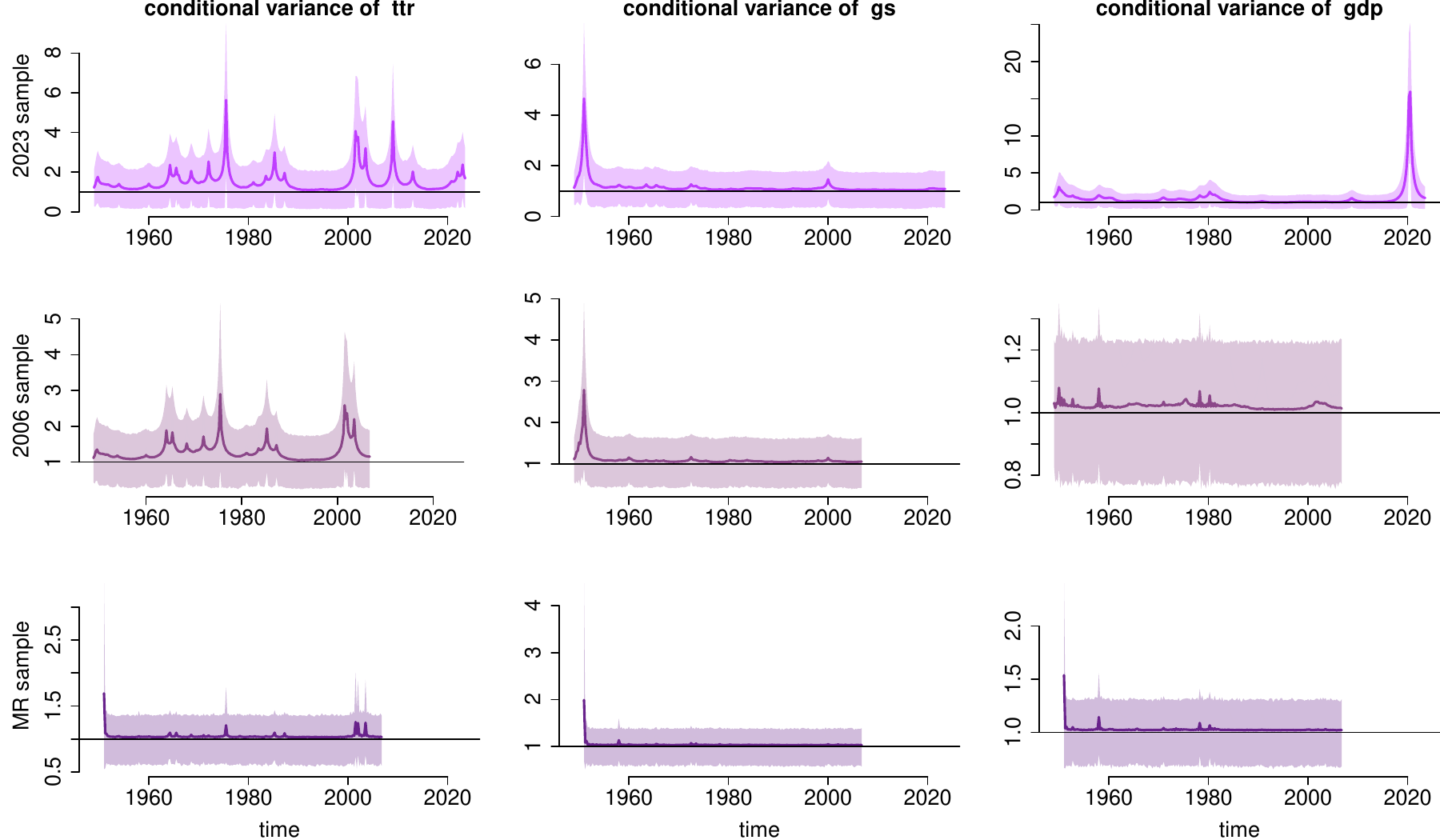}
\end{center}

\smallskip{\small
Note: The figures plot time-varying conditional variances of the structural shocks. The lines report the posterior mean and the shaded areas 90\% HPD intervals. The variances in the first row clearly exhibit non-proportional changes across time. The horizontal black line is set at the value of 1, around which the prior is centred.}
\label{fig:HV}
\end{figure}

The conditional variances in the 2006-sample are to
some extent similar
to those from the 2023-sample until 2006. However, at all times, the 90\% HPD intervals include
the value of~1. This is caused by a weaker signal provided from the data in the shorter sample regarding time-varying volatility, which undermines the evidence for identification
in the framework of our model. In the MR-sample,
the evidence for conditional variances that support
identification is even weaker. Thus, the bottom line is that,
in the 2023-sample, the shocks are clearly identified
through heteroskedasticity, while the evidence for
identification through heteroskedasticity is weaker in the
2006-sample, and no such evidence is found in the MR-sample.

\subsection{The effects of tax shocks}

\noindent Thus far, we document varying degrees of evidence of partial identification of structural shocks across samples. The MR-sample shows no evidence and is therefore excluded from further analysis. Given that heteroskedasticity provides three identified shocks for the 2023-sample, we examine which one corresponds to the tax shock. To do so, we begin by reporting the correlations between the structural shocks from our estimated model and the narrative measure of an unanticipated tax shock by \cite{romer_macroeconomic_2010}, based on the 2006- and 2023-samples. These correlations are presented in Table~\ref{tab:corr}.

\begin{table}[t]
\caption{Correlations between the narrative tax shock measure of \cite{romer_macroeconomic_2010} and the three structural shocks in our model, as well as with other tax shocks proposed in the literature}

\begin{center}
\begin{tabular}{crrrrr}
\toprule
Structural shocks & 2023-sample & 2006-sample & BP results & MR results & LE results \\
\midrule
$w_{t}^{ttr}$ & 0.224 & 0.264 & 0.277 & 0.298 & 0.233 \\
$w_{t}^{gs}$ & -0.022 & 0.030 &  &  &  \\
$w_{t}^{gdp}$ & -0.154 & -0.170 &  &  &  \\
\bottomrule
\end{tabular}
\end{center}

{\small
Note: The table reports sample correlations between the narrative measure of tax shocks proposed by \cite{romer_macroeconomic_2010} and used by MR. The results in the 2023-sample and 2006-sample columns are based on our posterior estimations, where we used the posterior mean of the shocks as their estimator. The results in the BP results, MR results, and LE results columns are based on our reproduction of the results from MR and LE using the authors' computer codes and data.
}
\label{tab:corr}
\end{table}

The results show that the first shock in our models is, albeit modestly, the most correlated with the narrative measure of \cite{romer_macroeconomic_2010}. Such low correlations are consistent with the \emph{weak instrument} observation of \cite{ramey2016macroeconomic}, yet exceed 0.22 for all models in Table~\ref{tab:corr}. They are also higher than those for the second shock, with values of -0.022 for the 2023-sample and 0.03 for the 2006-sample, and for the third shock, which reports correlations below -0.15 in both samples. Notably, the correlations for the first shock are similar to those of the tax shocks estimated by BP, MR, and LE, reported in the last three columns of Table~\ref{tab:corr}. Taken together, this correlation analysis provides some support for interpreting the first shock as the \emph{tax shock}.

\begin{figure}[t!]
 \caption{Impulse responses of gross domestic product to a negative tax shock: our estimates}

\begin{center}
\includegraphics[width=.8\textwidth]{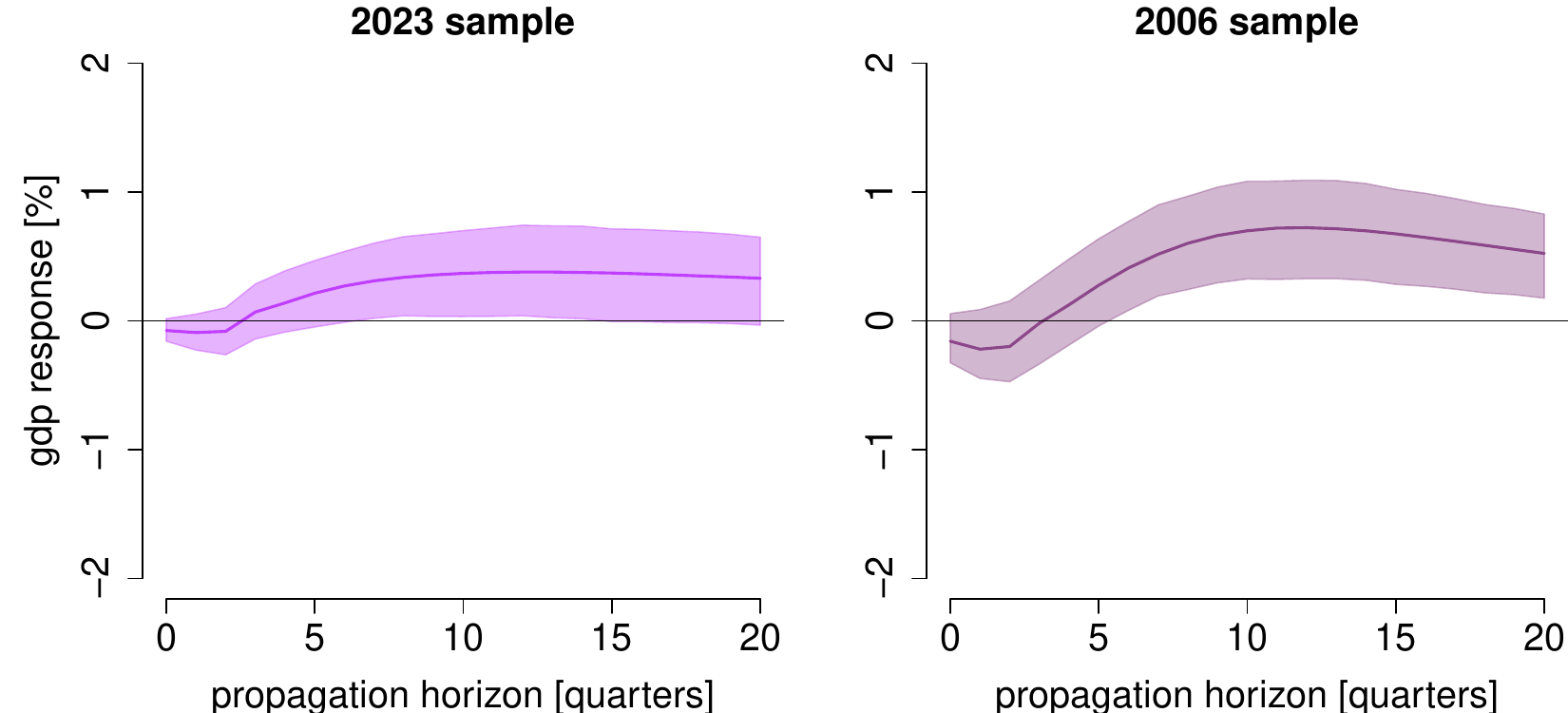}
\end{center}

\smallskip{\small
Note: The figure reports impulse responses of $gdp_t$ to a negative tax shock lowering $ttr_t$ by 1\% of the $gdp_t$ value in the last quarter of 2006. The lines report the posterior medians and the shaded areas the 68\% HPD point-wise intervals.
}
\label{fig:IRFown}
\end{figure}

Next, we investigate the dynamic effects of the tax shock identified with our approach on $gdp_t$. Figure~\ref{fig:IRFown} reports the corresponding impulse responses for both the 2006- and 2023-samples. Following MR and LE, these responses correspond to a tax shock that reduces $ttr_t$ by 1\% of $gdp_t$. The impulse responses in Figure~\ref{fig:IRFown} share two common features: (i) no effect on impact and during the first six quarters, and (ii) a subsequent increase in $gdp_t$, reaching a peak thirteen quarters after the shock at about 0.38\% for the 2023-sample and 0.73\% for the 2006-sample. The shorter-sample shock is more persistent, with effects that remain significant even five years after impact, whereas in the longer sample the response dies out after roughly 3.5 years. Overall, the shapes of the impulse responses from the 2023- and 2006-samples are quite similar.

\begin{figure}[t!]
 \caption{Impulse responses of gross domestic product to a negative tax shock: comparison with other studies}

\begin{center}
\includegraphics[width=.8\textwidth]{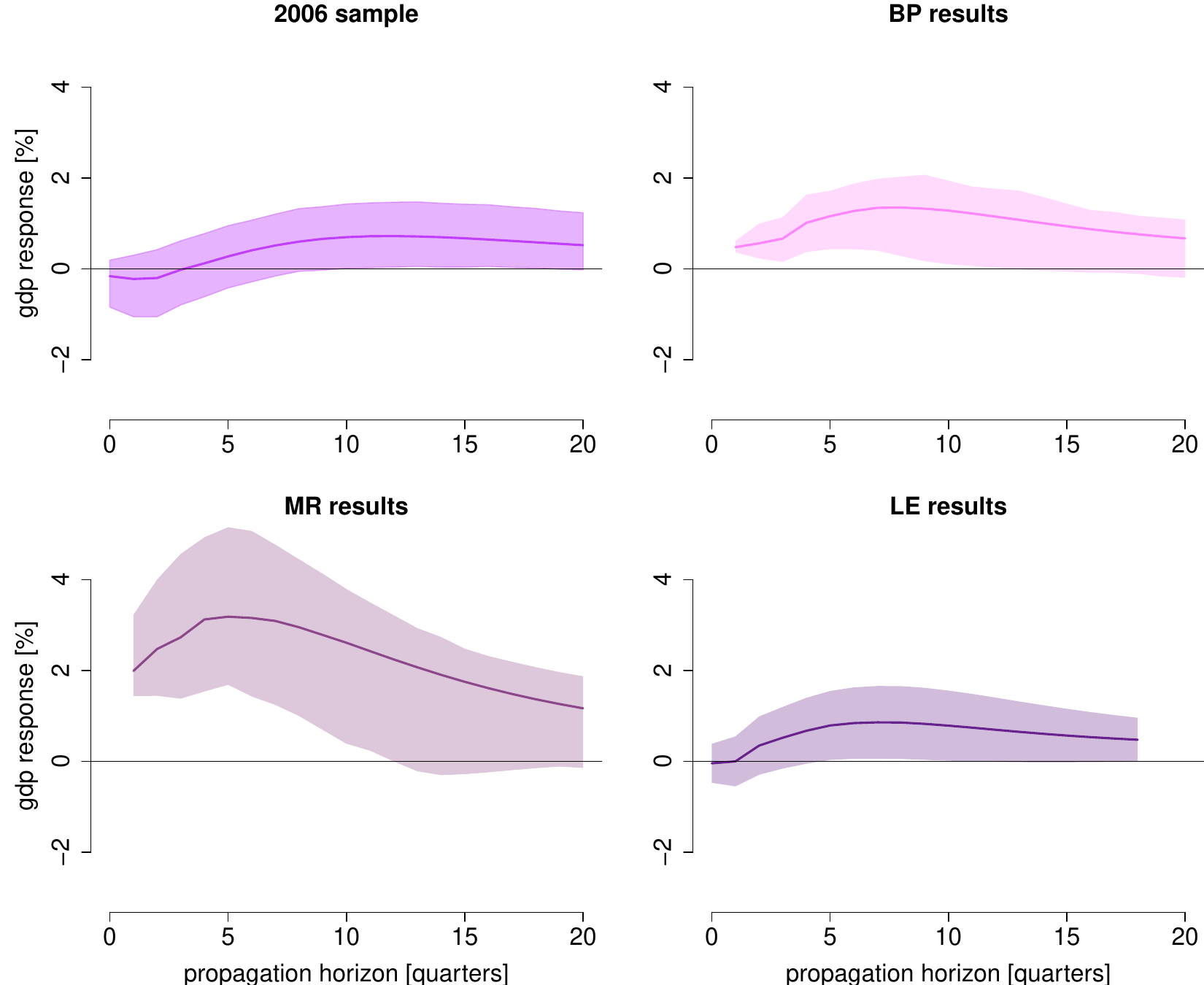}
\end{center}

\smallskip{\small
Note: The figure reports impulse responses of $gdp_t$ to a negative tax shock lowering $ttr_t$ by 1\% of the $gdp_t$ value in the last quarter of 2006. In the 2006-sample plot, the line reports the posterior median and the shaded area reports the 90\% HPD point-wise interval for the PM-ordering. In the remaining plots, the lines report the maximum likelihood estimator and the shaded areas, the 95\% point-wise confidence intervals.
}
\label{fig:IRFcom}
\end{figure}

We also examine how the impulse response based on our 2006-sample compares with those reported in BP, MR, and LE for the MR-sample, i.e., the original data used by these authors.\footnote{Our BP results rely on the SVAR model of \cite{blanchard_empirical_2002}, as estimated by \cite{mertens_reconciliation_2014}.} Figure~\ref{fig:IRFcom} presents these results with the 90\% HDP intervals for our model estimates. The impulse responses in BP, MR, and LE are maximum likelihood estimates with 95\% confidence intervals.  

Our results share two features with those in the literature: the peak occurs at mid-horizons, and statistical significance fades after three to four years. In addition, our peak response is close to those in BP and LE, whereas MR obtain a larger peak. A further difference is that only the impulse responses reported by LE are statistically insignificant on impact and in the following four quarters—as in our estimates—while BP and MR report positive and significant responses on impact. Overall, however, the conclusions from our estimates remain broadly consistent with the literature, even though they are obtained using a different identification approach, namely identification through non-centred stochastic volatility.

\section{Conclusion}\label{sec:Conclusion}

\noindent This paper studied how structural shocks can be identified in SVAR-SVs under Bayesian estimation, with a focus on the role of the non-centred parameterization for stochastic volatility. Our results highlight three main contributions.

First, we show that the non-centred parameterization leads to a marginal prior on the conditional variances of shocks that, unlike the common centred parameterization, is centred at homoskedasticity, with shrinkage and heavy tails. At the same time, the implied prior structure is consistent with the view—emphasized in earlier work on volatility comovement—that widespread heteroskedasticity in macroeconomic and financial series can often be traced back to a limited number of underlying shocks. In this sense, the non-centred parameterization not only facilitates partial identification, but also aligns the modeling framework with empirical evidence on the sources of volatility. As a byproduct of our main theoretical contribution, we also establish that the conditions for partial identification can be verified in practice. In particular, we derive the requirements under which such a verification procedure is feasible, and show how it can be implemented using the SDDR approach.

Second, our MC experiments demonstrate that adopting the non-centered parameterization yields systematic gains in estimation accuracy. Most notably, the parameters central to shock identification---the conditional variances and the corresponding rows of the impact matrix---exhibit consistently lower estimation error, with improvements in both small and large systems.

Third, applying our framework to well-known tax SVARs, we show that the non-centred approach can successfully identify tax shocks through heteroskedasticity. The resulting shocks align closely with estimates reported in the literature, indicating that the method is not only theoretically appealing but also reliable in applied settings.

Overall, the non-centred parameterization provides a simple but powerful way to 
exploit identification through stochastic volatility in SVARs. Although our analysis has focused on a fiscal 
application, the approach is easily adaptable to other areas of macroeconomics 
and finance where identification is partial or uncertain. An interesting avenue for 
future research is to examine its performance in more flexible settings, such as in 
\cite{CW2023}, where both the volatility of structural shocks and the impact matrix 
are allowed to vary over time.

\section{Acknowledgments}

\noindent For their useful comments and suggestions, we would like to thank Professor Subal Kumbhakar and Professor Sushanta Mallick (Editors), an anonymous Associate Editor, two anonymous referees, as well as participants at various seminars and conferences. This research did not receive any specific grant from funding agencies in the public, commercial, or not-for-profit sectors.

\bibliographystyle{chicago}
 \bibliography{SV_SVAR}

\appendix

\section{Partial identification in heteroskedastic SVARs: Theorem~A.1 and proofs}\label{app:proofs}

\noindent The following matrix result is the basis for
partial identification of structural parameters in 
structural VAR models with general heteroskedastic
or conditionally heteroskedastic 
structural shocks.

\begin{thm}\label{thm:identification}
Let $\mathbf{\Sigma}_t$, $t=0,1,\dots$, be a sequence of positive
definite $N \times N$ matrices and
$\boldsymbol{\Lambda}_t=\diag\left(\sigma^{2}_{1.t},\dots,\sigma^{2}_{N.t}\right)$
a sequence of $N\times N$ diagonal matrices
with $\boldsymbol{\Lambda}_0 = \mathbf{I}_N$.
Suppose there exists a nonsingular
$N\times N$ matrix $\mathbf{B}_0$ such that
\begin{equation}
\mathbf{\Sigma}_t=\mathbf{B}_0^{-1}\mathbf{\Lambda}_t \mathbf{B}_0^{-1\prime}, \quad t=0,1,\dots.
\label{e2}
\end{equation}
Let
$\boldsymbol{\sigma}^2_n=(1,\sigma^2_{n.1},\sigma^2_{n.2},\dots)$
be a possibly infinite dimensional vector.
Then the $n^{\rm th}$ row of $\mathbf{B}_0$
is unique up to sign if
$\boldsymbol{\sigma}^2_n\neq \boldsymbol{\sigma}^2_i$
\quad$\forall i \in \{1,\dots,N\}\backslash \{n\}$.
\end{thm}

Assuming that the $\mathbf{\Sigma}_t$ are the covariance matrices
of the reduced-form residuals in a VAR model and 
the $\mathbf{\Lambda}_t$ are the covariance matrices of the 
structural shocks, 
the theorem generalizes Theorem 1 of \cite{LW2017} which
presents an analogous result for structural errors with
volatility changes generated by a homogeneous Markov
switching process with finitely many volatility states. Our
Theorem~A.1
provides a~general result on identification of a single
equation through heteroskedasticity and also applies, for
instance, if the
volatility changes are generated by a different Markov
process for each shock. It shows that a
structural shock and, hence, the corresponding structural
equation is identified if the sequence of variances
is distinct from the
variance sequences of any of the other shocks.
Our Theorem~A.1
generalizes identification results for
some special volatility models that have been used in the
literature on identification through heteroskedasticity
\citep[see, e.g.,][Chapter 14]{KL:2017}. For example,
it is easy to see that identification results
for volatility models based on a finite number of
volatility regimes as considered by
\cite{Rigobon:03}, \cite{Rigobon/Sack:03},
\cite{Lanne/lue:06}, \cite{Lanne/lue/Ma:09},
\cite{Netsunajev:13}, \cite{Herwartz2011}, \cite{Wozniak2015},
 \cite{LV14}, \cite{luenet:14b}
follow directly from Theorem~A.1.

We emphasize that Theorem~A.1 also
implies identification conditions for 
the SV models considered in this study. In this context, 
SV models have also been proposed by 
\cite{lewis_identifying_2021} and \cite{Bertsche/Braun:18}. 
In such cases, the conditional covariance matrices of the 
reduced form errors are given by
$\mathbf{\Sigma}_t=\mathbf{B}_0^{-1}\mathbf{\Lambda}_t 
\mathbf{B}_0^{-1\prime}$,
where $\mathbf\Lambda_{t}=\diag\left(\sigma^2_{1.t}, \dots,
\sigma^2_{N.t} \right)$ is a diagonal matrix.
If the $\sigma^2_{n.t}$ vary stochastically, as in
SV dynamics, they will not be proportional with
probability 1 and, hence, satisfy the conditions for
identification of Theorem~A.1. So
if any one of the structural errors has changing
conditional variances, it will be identified, even
if all the other components have constant conditional
variance. That insight is used in our Bayesian analysis
of the SV model. It may be worth noting, however,
that Theorem~A.1 also implies that
a single shock may be homoskedastic and still be identified
in case all other shocks are heteroskedastic.
This discussion also shows that Theorem~A.1
generalizes results for full identification in
\cite{Sentana/Fiorentini:01}, \cite{lewis_identifying_2021},
and \cite{Bertsche/Braun:18} to the case of partial identification.

\subsubsection*{Proof of Theorem~A.1}

\noindent
We first prove the following lemma.

\begin{lm}\label{lm:identification}
Let $\mathbf{\Sigma}_t$, $t=0,1,\dots$, be a sequence of positive
definite $N \times N$ matrices and
$\boldsymbol{\Lambda}_t=\diag\left(\sigma^{2}_{1.t},\dots,\sigma^{2}_{N.t}\right)$
a sequence of $N\times N$ diagonal matrices
with $\boldsymbol{\Lambda}_0 = \mathbf{I}_N$.
Suppose there exists
a~nonsingular $N\times N$ matrix $\mathbf{B}$ such that
\begin{equation}\label{eq:lemma}
\mathbf{\Sigma}_t = \mathbf{B}\boldsymbol{\Lambda}_t\mathbf{B}', \qquad t=0,1,\dots.
\end{equation}
Let $\boldsymbol{\sigma}^2_n=(1,\sigma^2_{n.1},\sigma^2_{n.2},\dots)$
be a possibly infinite dimensional vector. Then the
$n^{\text{th}}$ column of $\mathbf{B}$ is unique up to sign if
$\boldsymbol{\sigma}^2_n\neq \boldsymbol{\sigma}^2_i$
\quad$\forall i \in \{1,\dots,N\}\backslash \{n\}$.
\end{lm}

\begin{proof}
Let $\mathbf{B}_*$ be a matrix that satisfies:
$\mathbf\Sigma_t = \mathbf{B}_*\mathbf\Lambda_t\mathbf{B}_*'$,
$t=0,1,\dots$.
It will be shown that, under the conditions of
Lemma \ref{lm:identification}, the $n^{\text{th}}$ column of
$\mathbf{B}_*$ must be the same as that of $\mathbf{B}$,
except perhaps for a reversal of signs. Without loss of
generality, it is assumed in the following that $n=1$ because
this simplifies the notation. In other words, it is shown that
the first columns of $\mathbf{B}$ and $\mathbf{B}_*$ are the
same except for a reversal of signs if
$\boldsymbol{\sigma}^2_1\neq \boldsymbol{\sigma}^2_i$,
$i=2,\dots,N$.

There exists a nonsingular $N\times N$ matrix $\mathbf{Q}$
such that $\mathbf{B}_* = \mathbf{BQ}$. Using
$\mathbf{\Sigma}_0 = \mathbf{B}\mathbf{B}'$, $\mathbf{Q}$
has to satisfy the relation
\begin{equation*}
\mathbf{BB}' = \mathbf{BQQ}'\mathbf{B}'.
\end{equation*}
Multiplying this relation from the left by $\mathbf{B}^{-1}$ and from the right by $\mathbf{B}^{-1'}$ implies that $\mathbf{QQ}' = \mathbf{I}_N$ and, hence, $\mathbf{Q}$ is an orthogonal matrix.

The relations
\begin{equation*}
\mathbf{B\Lambda}_t\mathbf{B}' = \mathbf{BQ\Lambda}_t\mathbf{Q}'\mathbf{B}'
\end{equation*}
imply $\mathbf\Lambda_t = \mathbf{Q\Lambda}_t\mathbf{Q}'$ and, hence, $\mathbf{Q\Lambda}_t = \mathbf{\Lambda}_t\mathbf{Q}$ for all $t=0,1,\dots$.

Denoting the $(i.j)^{\text{th}}$ element of $\mathbf{Q}$ by $q_{ij}$, the latter equation implies that
\begin{equation*}
q_{n1}\boldsymbol{\sigma}^2_1= q_{n1}\boldsymbol{\sigma}^2_n, \qquad n=1,\dots,N.
\end{equation*}
Hence, since $\boldsymbol{\sigma}^2_n$ is different from
$\boldsymbol{\sigma}^2_1$ for $n=2,\dots,N$, we must have 
$q_{n1} = 0$ for $n=2,\dots,N$. Since, $\mathbf{Q}$ is 
orthogonal, the first column must then be
\begin{equation*}
(1,0,\dots,0)' \qquad\text{or}\qquad (-1,0,\dots,0)'
\end{equation*}
which proves the lemma.
\end{proof}


Now consider the setup of Theorem~A.1 with
$\mathbf{B} = \mathbf{B}_0^{-1}$. Then the arguments in the
proof of Lemma \ref{lm:identification} show that 
$\mathbf{B}_{0*}^{-1} = \mathbf{B}_0^{-1}\mathbf{Q}$, 
where $\mathbf{Q}$ is as in the proof of Lemma 
\ref{lm:identification}. Hence, 
$\mathbf{B}_{0*} = \mathbf{Q}'\mathbf{B}_0$, 
which shows that $\mathbf{B}_{0*}$ and $\mathbf{B}_0$ have 
the same $n^{\text{th}}$ row up to sign. 
$\hspace*{\fill} \Box$\\

Theorem~A.1 implies that, if a structural
shock is identified, the corresponding structural impulse 
responses are unique. This result is stated formally in
the following corollary. 

\begin{cl}\label{cl2:IRF}
Under the conditions of Theorem~A.1,
if the $n^{\text{th}}$ row of $\mathbf{B}_0$ is unique, then 
for a given sequence of $N\times N$ matrices
$\boldsymbol\Phi_i$, $i=0,1,\dots$, the $n^{\rm th}$ column 
of the matrix
$\boldsymbol\Theta_i = \boldsymbol\Phi_i\mathbf{B}_0^{-1}$,
is unique for $i=0,1,\dots$.
\end{cl}

\begin{proof}
It has to be shown that uniqueness of the $n^{\text{th}}$ row of 
$\mathbf{B}_0$ implies a unique $n^{\text{th}}$ column of 
$\mathbf{B}_0^{-1}$. Without loss of generality we focus on 
the first row. If the first row of $\mathbf{B}_0$ is unique, 
any other admissible $\mathbf{B}_0$ matrix must be of the form 
$\mathbf{QB}_0$, where $\mathbf{Q}$ is an orthogonal matrix of 
the form:
\begin{equation*}
\begin{bmatrix} 1 & \mathbf{0}_{(1\times(N-1))} \\ 
\mathbf{0}_{((N-1)\times 1)} & \mathbf{Q}_*\end{bmatrix},
\end{equation*}
with $\mathbf{Q}_*$ being an orthogonal $(N-1) \times (N-1)$ 
matrix. This fact follows from the arguments in 
the proof of Lemma \ref{lm:identification}. Thus, any admissible 
inverse has the form $\mathbf{B}_0^{-1}Q'$ and, hence, has 
the same first column as $\mathbf{B}_0^{-1}$. Clearly, the 
same argument applies for any other row of $\mathbf{B}_0$. 
\end{proof}

\includepdf[pages=-]{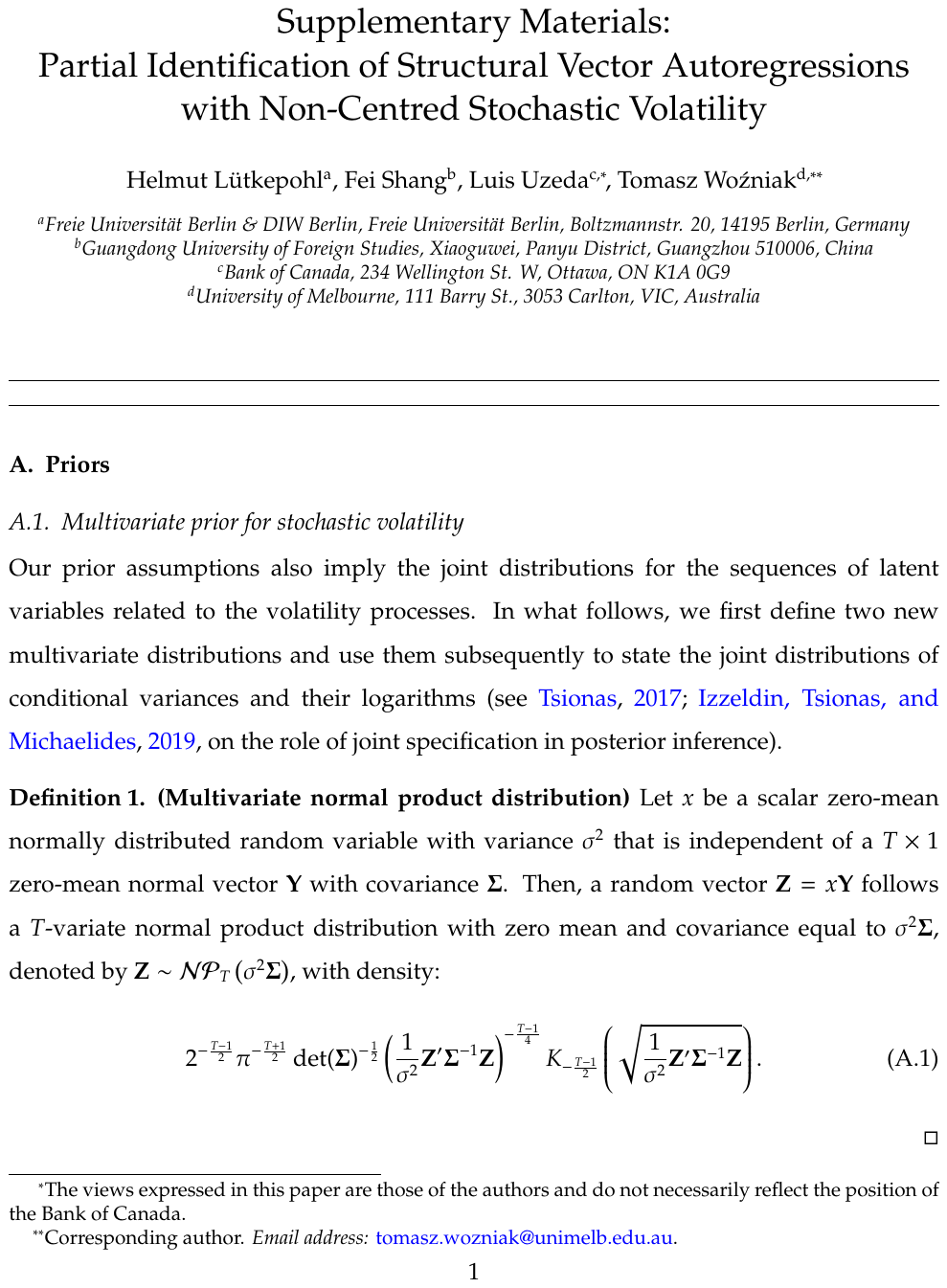}

\end{document}